\documentclass{aa}

\usepackage{graphicx}
\usepackage{txfonts}
\usepackage{natbib}
\usepackage{textcomp}
\usepackage{ulem}
\usepackage[colorlinks=true, citecolor=blue]{hyperref}
\usepackage{caption}
\usepackage{subcaption}
\usepackage{soul}
\def\kms{\hbox{km$\;$s$^{-1}$}\,}

\def\mAA{m\AA\,}

\def\FeIline{\ion{Fe}{i}~6173\,\AA\,}
\def\CaIR{\ion{Ca}{ii}~8542\,\AA\,}
\def\CaK{\ion{Ca}{ii}~K\,}
\def\CaH{\ion{Ca}{ii}~H\,}

\def\CaH{\ion{Ca}{ii}~H\,}
\def\CaHK{\ion{Ca}{ii}~H\&K\,}
\def\SiIVline{\ion{Si}{iv}~1403\,\AA\,}
\def\Halpha{\mbox{H\hspace{0.1ex}$\alpha$}\,}
\def\logtau{$\log\tau_{\mathrm{500}}$\,}
\def\lambdat{$\lambda$--$t$\,}
\def\blos{$B_{\mathrm{LOS}}$\,}
\def\roi{ROI\,}
\def\konev{$\mathrm{K_{1V}}$\,}
\def\ktwov{$\mathrm{K_{2V}}$\,}
\def\ktwor{$\mathrm{K_{2R}}$\,}
\def\kone{$\mathrm{K_{1}}$\,}
\def\ktwo{$\mathrm{K_{2}}$\,}
\def\kthree{$\mathrm{K_{3}}$\,}
\def\temp{$T$\,}
\def\vlos{$V_{\mathrm{LOS}}$\,}
\def\vturb{$V_{\mathrm{turb}}$\,}

\begin{document}

   \title{Properties of shock waves in the quiet Sun chromosphere}

   \author{Harsh Mathur
          \inst{1}
          \and
          Jayant Joshi
          \inst{1}
          \and
          K.Nagaraju\inst{1}
          \and
          Luc Rouppe van der Voort
          \inst{2, 3}
          \and
          Souvik Bose
          \inst{4,5,2,3}
          }

   \institute{Indian Institute of Astrophysics, 
              II Block, Koramangala, Bengaluru 560 034, India\\
              \email{harsh.mathur@iiap.res.in}
             \and
             Institute of Theoretical Astrophysics, University of Oslo, 
             P.O. Box 1029 Blindern, NO-0315 Oslo, Norway
             \and
             Rosseland Centre for Solar Physics, University of Oslo, 
             P.O. Box 1029 Blindern, NO-0315 Oslo, Norway
             \and
             Lockheed Martin Solar \& Astrophysics Laboratory, Palo Alto, CA 94304, USA
             \and
             Bay Area Environmental Research Institute, NASA Research Park, Moffett Field, CA 94035, USA
             }

   \date{Received 23.06.2022; accepted 03.10.2022}

 
  \abstract
   {Short-lived (100s or less), sub-arcsec to a couple of arcsec size features of enhanced brightenings in the narrowband images at the $\mathrm{H_{2V}}$ and \ktwov{} positions of the \CaHK lines in the quiet Sun are known as bright grains.
   These bright grains are interpreted as manifestations of acoustic shock waves in the chromosphere.}
   {We aim to study time-varying stratified atmospheric properties, such as the temperature, line-of-sight (LOS) velocity, and microturbulence inferred from observations of the bright grains during such acoustic shock events.}
   {With simultaneous observations of a quiet Sun \mbox{internetwork} region in the \FeIline, \CaIR, and \CaK lines acquired by the CRisp Imaging Spectro-Polarimeter and the CHROMospheric Imaging Spectrometer instruments on the Swedish 1-m Solar Telescope, we performed multi-line non-local thermodynamic equilibrium inversions using the STockholm inversion Code to infer the time-varying stratified atmosphere’s physical properties.}
   {The \CaK profiles of bright grains show enhancement in the \ktwov{} peak intensities with absence of the \ktwor{} features.
   At the time of maximum enhancement in the \ktwov{} peak intensities, we found average enhancements in temperature at lower chromospheric layers (at \logtau{} $\simeq-$4.2) of about 1.1~kK with a maximum enhancement of $\sim4.5$~kK.
   These temperature enhancements are colocated with upflows, as strong as $-$6~\kms{}, in the direction of the LOS.
   The LOS velocities at upper chromospheric layers at \logtau{}$<-$4.2 show consistent downflows greater than $+$8~\kms{}.
   The retrieved value of microturbulence in the atmosphere of bright grains is negligible at chromospheric layers.  
   }
   {
   This study provides observational evidence to support the interpretation that the bright grains observed in narrowband images at the $\mathrm{H_{2V}}$ and \ktwov{} positions of the \CaHK{} lines are manifestations of upward propagating acoustic shocks against a background of downflowing atmospheres.
   }


   \maketitle
%

\section{Introduction}
\label{sect:introduction}
Waves are omnipresent in the solar atmosphere.
Understanding the source, nature, and energy dissipation mechanism of waves in the solar atmosphere is essential to know possible contributions to chromospheric and coronal heating.

Narrowband images at the $\mathrm{H_{2V}}$ and \ktwov{} positions of the \CaHK{} lines in the \mbox{internetwork} region of chromosphere show short-lived (100~s or less), sub-arcsec to a couple of arcsec size brightenings called grains \citep{1964PhDT........83B}.
These bright grains have also been observed in the Transition Region and Coronal Explorer near-UV passbands \citep{1999ASPC..183..383R, 1999SoPh..190..351H, 2010A&A...519A..58T}.
The typical size of these grains in the \CaK{} filtergrams observed with the German Vacuum Tower Telescope \citep{1998NewAR..42..493V} and narrowband imaging spectroscopy in the \CaIR{} line from the Dunn Solar Telescope \citep[DST;][]{1969S&T....38..368D} are of the order of 1\farcs95 \citep{2006A&A...459L...9W} and 0\farcs8 \citep{2009A&A...494..269V}, respectively.

Decades of observations have suggested these to be manifestations of acoustic shock like disturbances that propagate upward in the solar atmosphere \citep{1974SoPh...37...75C, 1974SoPh...38...91Z, 1976A&A....50..263C, 1977A&A....57..211C, 1997ApJ...481..500C, 2008A&A...479..213B}.
These grains are not known to be caused by magnetism-related processes \citep{1990IAUS..142..197K, 1991SoPh..134...15R, 1999ApJ...517.1013L}.
We refer the reader to the \cite{1991SoPh..134...15R} (and the references therein) for an extensive review on early efforts of observation, simulation, and interpretation of grains.

\citet{1992ApJ...397L..59C, 1997ApJ...481..500C} reproduced the bright grains in the \CaK line using one-dimensional hydrodynamic models with radiation in non-local thermodynamic equilibrium (non-LTE) assuming complete redistribution (CRD).
Quiet-Sun acoustic shock waves have been studied by \citet{2004A&A...414.1121W}  using three-dimensional simulation assuming LTE conditions.
These simulations showed that upward propagating acoustic waves from the lower atmosphere turn into shocks in the chromosphere due to a drop in the gas density by several orders of magnitude \citep{1992ApJ...397L..59C, 2004A&A...414.1121W, 2014ApJ...784...20D}. 

The density decreases while the shock continues to travel upward from the chromosphere to the transition region, leading to steepening of the shock and an increase in the shock amplitude.
As a result, one would expect to see a transition region counterpart of the bright grains.
However, observational studies from 90’s \citep{1997ApJ...490L.195J, 1997ESASP.404..685S} from the Solar Ultraviolet Measurements of Emitted Radiation \citep[SUMER,][]{1995SoPh..162..189W} instrument on board the Solar and Heliospheric Observatory \citep{1995SoPh..162....1D} suggested that the bright grains do not have a transition region counterpart.
Thus, it lead to the conclusion that upward propagating acoustic shock waves do not play a significant role in the heating of the transition region \citep{1997ApJ...490L.195J}.
More recently, however, using observations from the Interface Region Imaging Spectrograph \citep[IRIS,][]{2014SoPh..289.2733D} and the Swedish 1-m Solar Telescope \citep[SST,][]{2003SPIE.4853..341S},  \citet{2015ApJ...803...44M} showed that some bright grains have signatures in the \SiIVline line, suggesting that some acoustic shock waves can in fact reach the transition region.
%
Using imaging spectroscopic observations of the quiet internetwork chromosphere in the \CaIR{} line, \citet{2009A&A...494..269V} suggested that some acoustic waves may be unable to propagate to the transition region because horizontal magnetic field lines (the canopy) prevent acoustic shock waves from propagating to the outer solar atmosphere.

Semi-empirical models and magneto-hydrodynamics (MHD) simulations have also suggested that upward propagating acoustic shocks cause brightenings in the millimeter continuum \citep{2006IAUS..233..104L, 2007A&A...471..977W, 2020A&A...634A..56D, 2021RSPTA.37900185E}.
Through analysis of data from the Atacama Large Millimeter Array \citep[ALMA;][]{5136193}, \citet{2020A&A...644A.152E} found a typical chromospheric temperature enhancement of 400--750~K during the shock with a maximum enhancement of 1200~K.
Moreover, \citet{2021A&A...656A..68E} suggested that the ALMA observations studied by \citet{2020A&A...644A.152E} had limited spatial resolution that can lead to reduced brightness temperatures and underestimation of the temperature enhancement.

The spectra of grains in the \CaH{} line recorded by the POlarimetric LIttrow Spectrograph \citep[POLIS,][]{2005A&A...437.1159B} was inverted by \citet{2013A&A...553A..73B} using the LTE approximation.
They found an increase in temperature of 200--300~K in the quiet Sun chromosphere.

%
%
The limitations in these earlier observational studies, like the limited spatial resolution of ALMA and the LTE approximation used in inversions of the \CaH{} spectra merits further observational work with advanced methods.
In this study, we use high-resolution imaging-spectropolarimetric observations of the bright grains in the \CaK{} line from the Swedish Solar Telescope  to infer the time-varying stratified atmospheric properties, such as the temperature, Doppler shift, and microturbulence using a non-LTE inversion code.

\section{Observations}
\label{sect:observations}

\begin{figure*}[htbp]
\centering
            \includegraphics[width=0.88\textwidth]{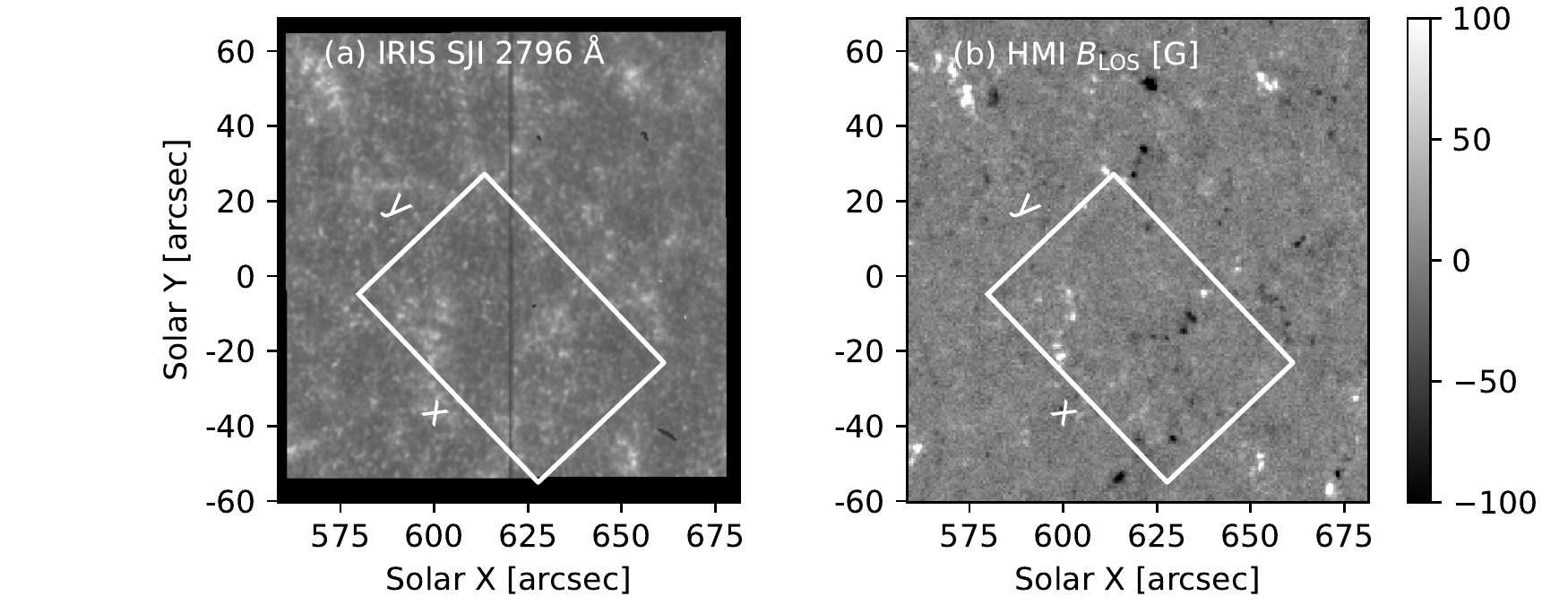}
            \includegraphics[width=0.86\textwidth]{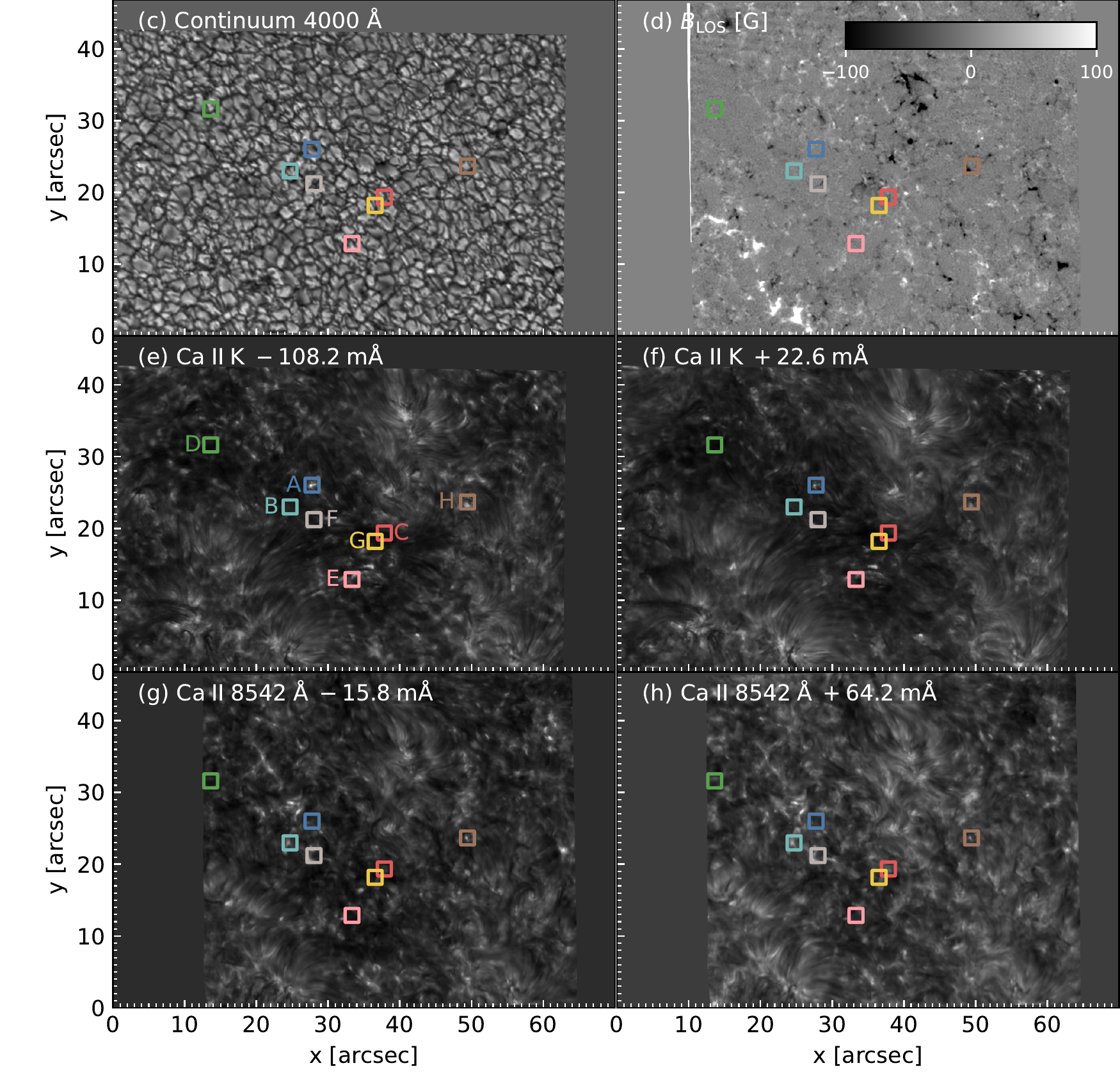}
      \caption{Overview of quiet Sun region observed on 6 June 2019. (a) IRIS slit-jaw image (SJI) at the \ion{Mg}{II}~k 2796~\AA\,. (b) SDO / HMI line-of-sight (LOS) magnetogram. The white box in panels (a) and (b) marks the spatial location of the FOV observed with the CHROMIS instrument. (c) Continuum image at 4000~\AA. (d) LOS magnetic field inferred using Milne-Eddington inversions of the \FeIline line observed with the CRISP instrument. Panels (e) and (f) show images at wavelength offsets of $-108.2$ and $+22.6$~\mAA from the \CaK{} line core, respectively. Panels (g) and (h) display images at wavelength offsets of $-15.8$ and $+64.2$~\mAA from the \CaIR{} line core, respectively. A gamma correction with $\gamma=1.5$ and $\gamma=0.5$ has been applied on panels (a) and (e)--(h), respectively. The regions of interest (ROI, namely A--H) highlighted by colored squares of $50 \times 50$ pixels ($1\farcs85 \times $1\farcs85) indicate locations of \CaK{} grains that are analyzed in this paper. The \roi{A} shows a CBG in \CaK{} narrowband images (panels (e) and (f)), the brightenings in other ROIs occured at different times.
             }
         \label{fig:FOV}
\end{figure*}

\begin{figure*}
\begin{center}
    \textbf{ROI A}
\end{center}
\begin{subfigure}{0.5\textwidth}
\resizebox{\textwidth}{!}{\includegraphics{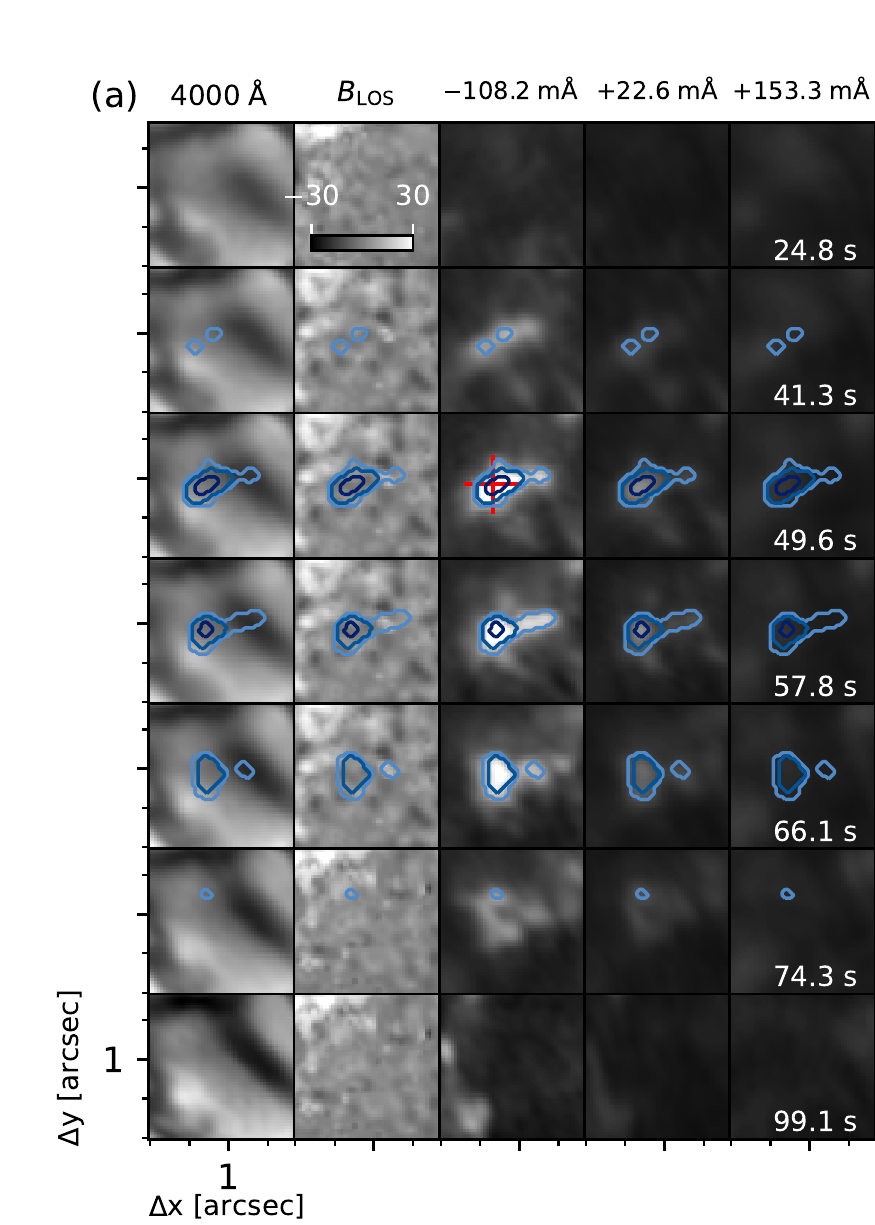}}
\end{subfigure}
\begin{subfigure}{0.5\textwidth}
\begin{subfigure}{0.495\textwidth}
\resizebox{\textwidth}{!}{\includegraphics{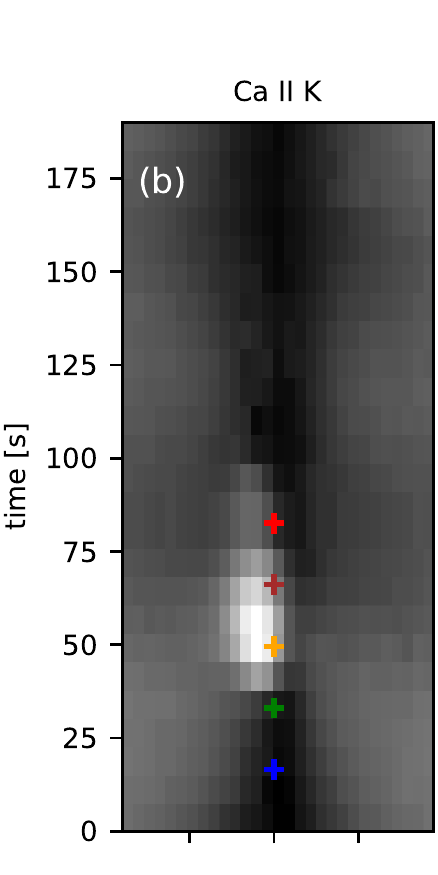}}
\resizebox{\textwidth}{!}{\includegraphics{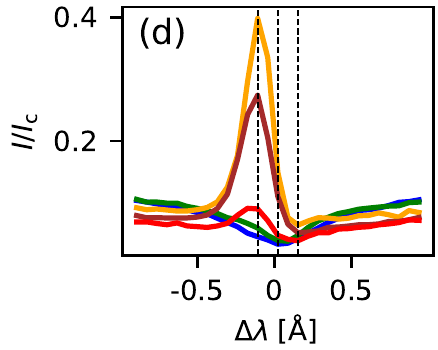}}
\end{subfigure}
\begin{subfigure}{0.495\textwidth}
\resizebox{\textwidth}{!}{\includegraphics{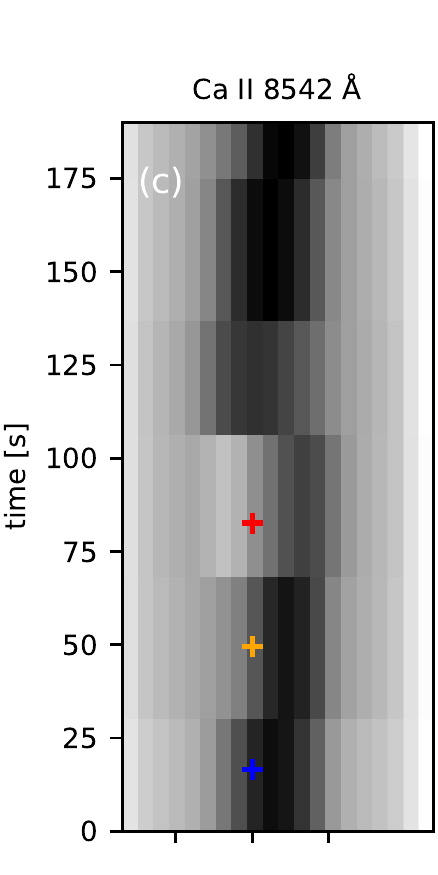}}
\resizebox{\textwidth}{!}{\includegraphics{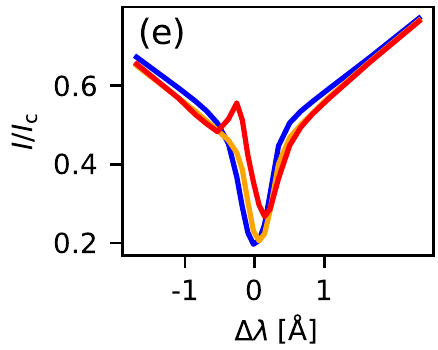}}
\end{subfigure}
\end{subfigure}
\caption{\label{fig:shocksevolution1}
Region of Interest (ROI) A: (a) The time evolution of the continuum at 4000~\AA, line of sight magnetic field (\blos{}), and images at wavelength offsets of $-108.2$~\mAA, $+22.6$~\mAA and $+153.3$~\mAA from the core of the \CaK{} line are shown column-wise. Panels (b) and (c) display \lambdat diagram for the pixel marked with '+' in panel (a) in the \CaK and \CaIR lines, respectively. A gamma correction with $\gamma=0.85$ is applied on the \CaK{} narrowband images in panel (a) and $\gamma=0.1$ on panels (b) and (c). A few selected profiles marked in panels (b) and (c) are shown in panels (d) and (e). The outer contour (cyan) is made with pixels belonging to RPs 85, 36, 18, and 78, while the middle contour (azure blue) is made with RPs 18 and 78 in panel~(a). The inner contour (navy blue) shows the region belonging to RP 78, the strongest CBG RP. The dashed vertical lines in panel (d) show the position of narrowband images in panel (a).
}
\end{figure*}

\begin{figure*}[htbp]
\begin{center}
    \textbf{ROI B}
\end{center}
\begin{subfigure}{0.5\textwidth}
\resizebox{\textwidth}{!}{\includegraphics{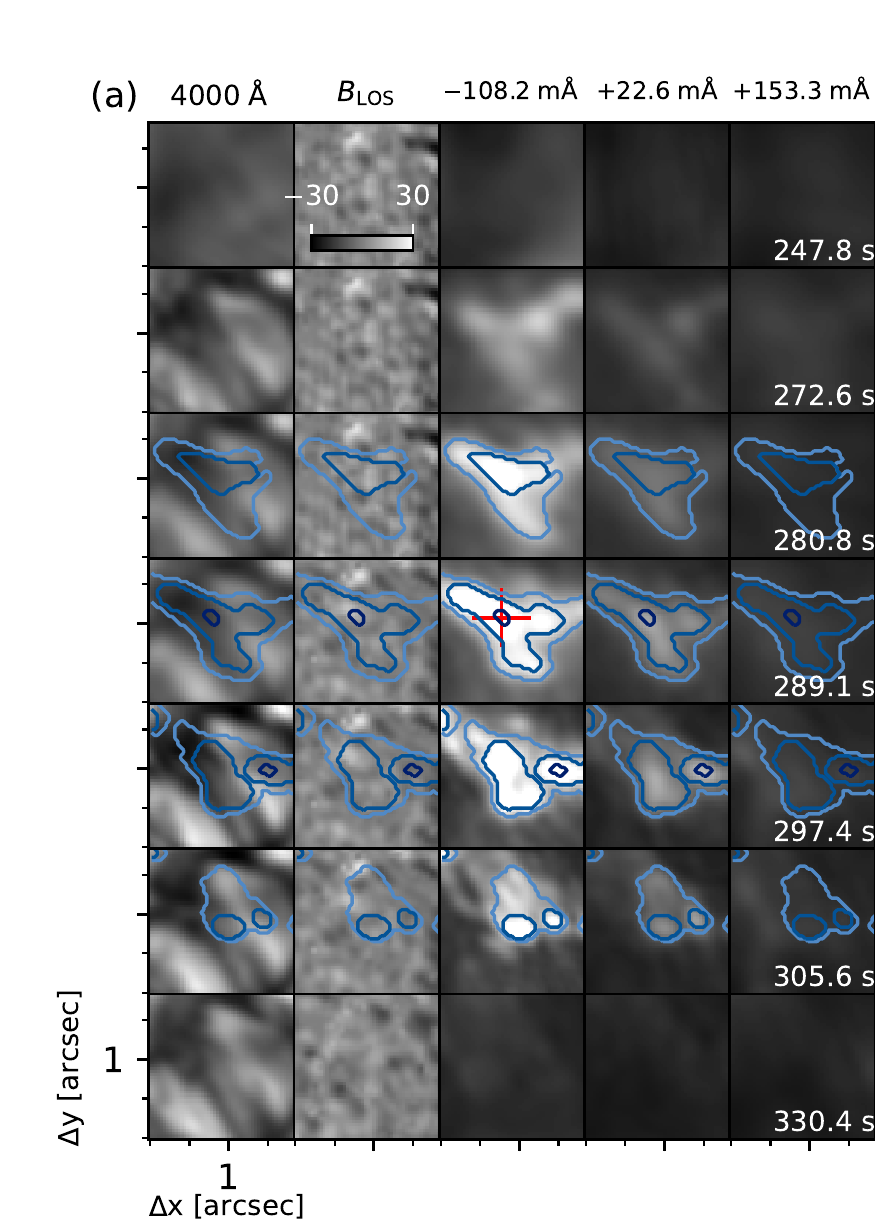}}
\end{subfigure}
\begin{subfigure}{0.5\textwidth}
\begin{subfigure}{0.495\textwidth}
\resizebox{\textwidth}{!}{\includegraphics{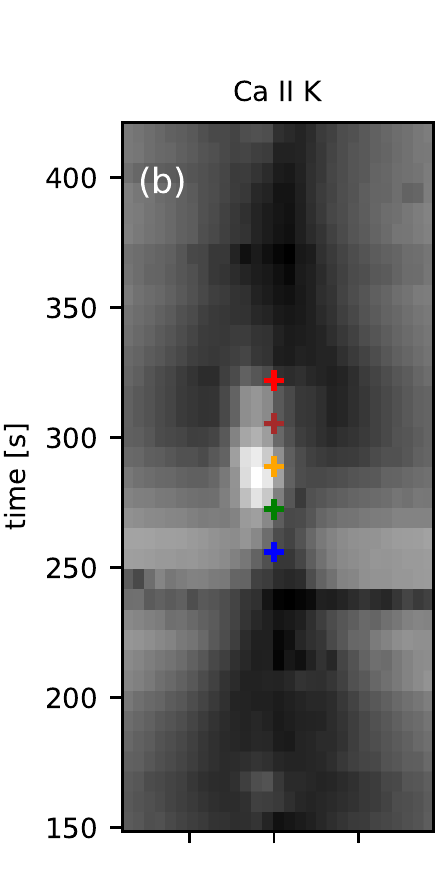}}
\resizebox{\textwidth}{!}{\includegraphics{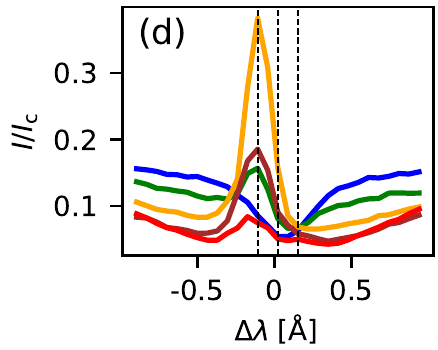}}
\end{subfigure}
\begin{subfigure}{0.495\textwidth}
\resizebox{\textwidth}{!}{\includegraphics{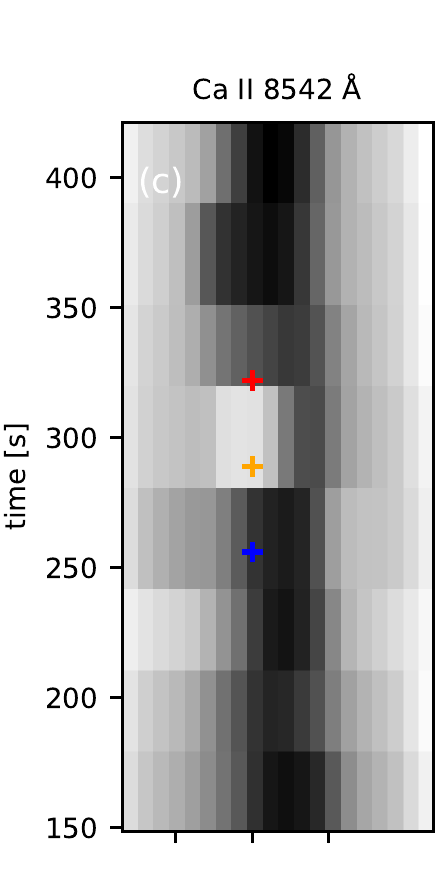}}
\resizebox{\textwidth}{!}{\includegraphics{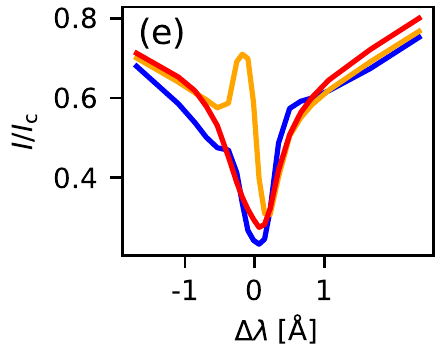}}
\end{subfigure}
\end{subfigure}
\caption{\label{fig:shocksevolution2}
same as figure \ref{fig:shocksevolution1} for \roi{B}.
}
\end{figure*}

\begin{figure*}[htbp]
\begin{center}
    \textbf{ROI C}
\end{center}
\begin{subfigure}{0.5\textwidth}
\resizebox{\textwidth}{!}{\includegraphics{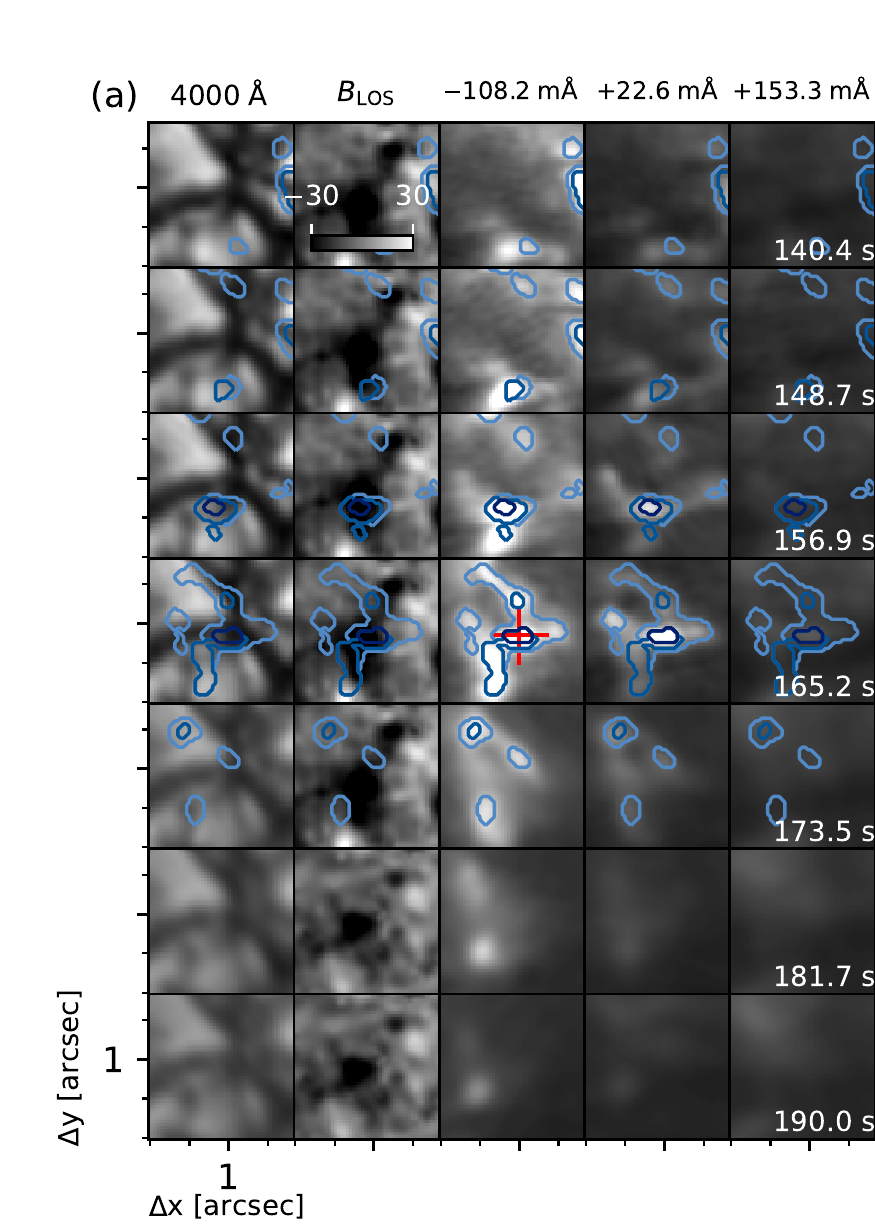}}
\end{subfigure}
\begin{subfigure}{0.5\textwidth}
\begin{subfigure}{0.495\textwidth}
\resizebox{\textwidth}{!}{\includegraphics{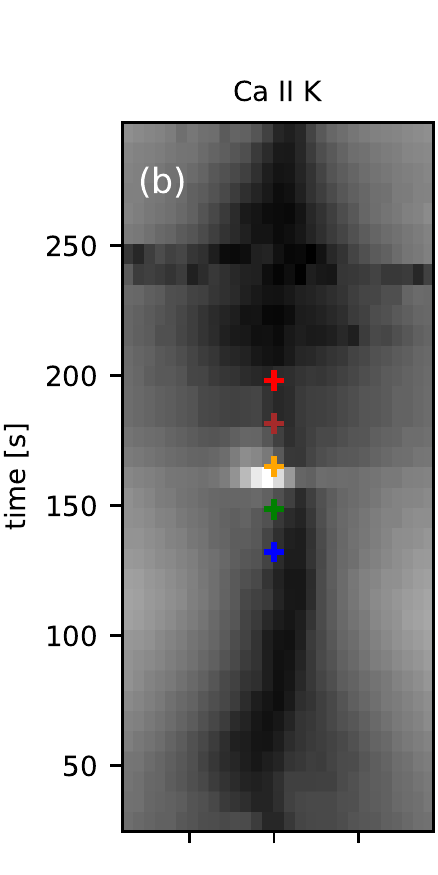}}
\resizebox{\textwidth}{!}{\includegraphics{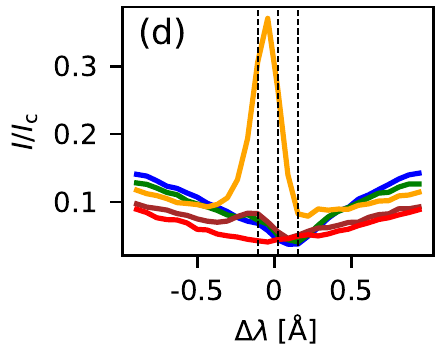}}
\end{subfigure}
\begin{subfigure}{0.495\textwidth}
\resizebox{\textwidth}{!}{\includegraphics{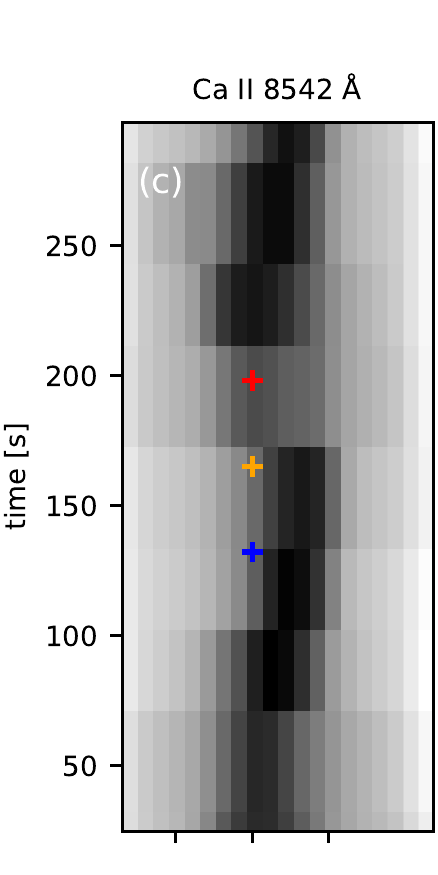}}
\resizebox{\textwidth}{!}{\includegraphics{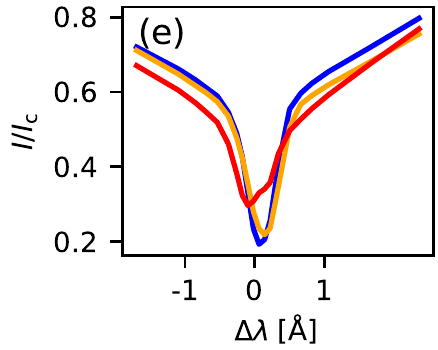}}
\end{subfigure}
\end{subfigure}
\caption{\label{fig:shocksevolution3}
same as figure \ref{fig:shocksevolution1} for \roi{C}.
}
\end{figure*}

    The observations were made with SST on 2019-06-06 between 10:26:20--10:40:06 UT.
The observed field-of-view (FOV) consist of a quiet Sun scene centered at solar $(x, y) = (631 \arcsec, -23 \arcsec)$, corresponding to a viewing angle $\cos\theta = \mu = 0.7$. Here, $\theta$ represents the angle with respect to the solar normal.
The observations were obtained with the CRisp Imaging Spectro-Polarimeter \citep[CRISP;][]{2008ApJ...689L..69S} and CHROMospheric Imaging Spectrometer \citep[CHROMIS;][]{2017psio.confE..85S} instruments simultaneously.
The observed FOV of CHROMIS covers an area of $50\arcsec \times 33\arcsec $, which overlaps with the CRISP FOV of $41\arcsec \times 37\arcsec$ as shown in Fig.~\ref{fig:FOV}.

The CRISP instrument sampled the \CaIR line with 20 non-equidistant wavelength points spanning from $-1.69$~\AA\, to $+2.36$~\AA\, and the \FeIline line was observed in 14 non-equidistant wavelength points between $-0.35$~\AA\, and $+0.64$~\AA.
We acquired observations in these lines with full-Stokes polarimetry.
The pixel scale of the CRISP data are 0\farcs059 and the observation sequences have a cadence of 37~s.

With CHROMIS, we observed the \CaK line in 29 wavelength positions around the line center (3933.682~\AA), with 65 \mAA sampling.
CHROMIS also observed one additional wavelength position in the continuum at 4000~\AA.
The data from CHROMIS instrument have a cadence of 8.26~s and a pixel scale of 0\farcs038.

Wide-band (WB) images using an auxiliary channel with a filter centered at 3950~\AA\, (FWHM = 6.5~\AA) and a filter centered at the \Halpha{} line (FWHM = 4.9~\AA) are also observed using the CHROMIS and CRISP instruments, respectively.
The details of the optical setup along with the passbands of the different filters are given in \citet{2021A&A...653A..68L}.

High-quality spatial resolution is achieved by the combination of good seeing conditions, the adaptive optics system, and the excellent CRISP and CHROMIS re-imaging systems \citep{2003SPIE.4853..370S, 2019A&A...626A..55S}.
The raw data were then processed using the CRISPRED \citep{2015A&A...573A..40D} and CHROMISRED \citep{2021A&A...653A..68L} pipelines for CRISP and CHROMIS data, respectively.
The reduction process includes image restoration by using Multi-Object Multi-Frame Blind Deconvolution \citep[MOMFBD;][]{2005SoPh..228..191V}.
The images from both the instruments were de-rotated to account for diurnal field rotation, aligned and de-stretched to remove warping due to seeing effects before making it ready for scientific analyses.
The CRISP data were co-aligned with CHROMIS data using cross-correlation of photospheric WB channels.
Before the alignment, the CRISP data were resampled to the CHROMIS pixel scale, and its cadence were adjusted to CHROMIS cadence by using nearest-neighbor interpolation.
The last step in the post-process procedure was to perform absolute intensity calibration of the data by comparing average line profiles observed at the disk center with the Hamburg atlas spectrum \citep{1984SoPh...90..205N}.

We also performed Milne-Eddington (ME) inversions of the \FeIline\ data to infer the line of sight magnetic field (\blos{}) utilizing the pyMilne code, a parallel C++/Python implementation\footnote{\url{https://github.com/jaimedelacruz/pyMilne}} \citep{2019A&A...631A.153D}.

Figure~\ref{fig:FOV} shows an overview of the observations at one time step.
The co-temporal IRIS slit-jaw image (SJI) in the \ion{Mg}{II}~k~2796~\AA\,, and the Solar Dynamics Observatory (SDO) / Helioseismic and Magnetic Imager (HMI) magnetogram serves as a reference which shows an extended quiet Sun scene, with enhanced emission in the IRIS SJI associated with the magnetic network elements seen in the HMI magnetogram (see panels (a) and (b)).
The white boxes in panels (a) and (b) mark the spatial location of the FOV observed through the CHROMIS instrument.
The continuum image (see panel (c)) displays regular granulation in the quiet Sun. 
The \blos map, inferred from ME inversions of the \FeIline{} line, (see panel (d)) shows magnetic network regions around $(x,y)$ = (20\arcsec, 10\arcsec) and $(x,y)$ = (45\arcsec, 35\arcsec), whereas, the central part of the FOV corresponds to an \mbox{internetwork} region.
Panel~(e) of Fig.~\ref{fig:FOV} shows transient brightenings in the \CaK{} blue wing image (at $-108.2$~m\AA).
A hint of these transient brightenings can also be seen in narrowband images near to the line core of the \CaK{} line (at wavelength offset $+$22.6~\mAA{}, see panel (f)).
For example, there is a brightening in the blue colored square (\roi{A}), the brightening events at other colored squares occurred at different times.
Signatures of these transient brightenings are visible in the blue wing images of the \CaIR line (at $-15.8$~\mAA, see panel (g)) but not in the red wing images (at $+$64.2~\mAA, see panel (h)).
These compact brightenings are prevalent in the \mbox{internetwork} region and their typical lifetime is found to be shorter than 60~s. 
These small scale and short-lived brightenings in the \CaK narrowband images are referred to as chromospheric bright grains (CBGs), which are the focus of this study.

\section{Methods of analysis}
\label{sect:methodsofanalysis}

\subsection{Spatial and spectral signatures of the CBGs}
\label{subsect:spatialspectral}

We have selected eight small regions of interest (\roi{}, namely A--H) of $50$ $\times$ $50$ pixels covering an area of $1\farcs85$ $\times$ $1\farcs85$.
Depending on the duration of the CBG events, we selected a minimum of 7 and a maximum of 12 time step sequences for each \roi{}, adding up to 78 time steps for all \roi{}s.
The selected \roi{}s are highlighted by colored squares in Fig.~\ref{fig:FOV}.
The lifetime of the presented CBGs ranges between $25$~s and $67$~s.

Figure~\ref{fig:shocksevolution1} shows the evolution of the CBG in \roi{A}.
At $t=41.3$~s, a CBG appears in the blue wing image of the \CaK line (at $-108.2$~\mAA).
The contour(s) of the CBG are created using a few profiles that are representative of the CBG, Representative Profiles (RPs), discussed in section~\ref{subsect:classificationdataset}.
The CBG grows in size and increases in intensity till $t=57.8$~s and then starts declining in size and vanishes completely after 74.3~s.
Signatures of the CBG are not visible in the \CaK line red wing (at $+153.3$~\mAA) images, whereas there is a very weak intensity enhancement in the \CaK images near to the line core ($+22.6$~\mAA)  at $t= $ 49.6~s and 57.8~s.
The corresponding continuum images at 4000~\AA\, display a typical granulation pattern and the \blos maps indicate minimal magnetic activity with a maximum photospheric field strength of $\sim20$~G.
The wavelength-time (\lambdat) diagram in panel~(b) for the brightest pixel in the CBG shows the evolution of short-lived intensity enhancement in the blue wing of the \CaK{} line.
Selected \CaK profiles are shown in panel~(d) at different stages during the lifetime of the CBG.
Prior to the onset of CBG, the \kone{} and the \ktwo{} features are absent and only the \kthree{} absorption is present.
The nomenclature of the \kone{}, \ktwo{} and \kthree{} spectral features in the \CaK{} line are described in Fig.~1 of \cite{1991SoPh..134...15R}.
At the onset of the CBG, we see an enhancement in the intensity of the \ktwov{} peak, which increases as the CBG reaches its maximum with the \kthree{} feature redshifted ($\Delta\lambda=+$153~\mAA{}).
At the end of the CBG, the \CaK{} profiles appear similar to the quiescent absorption profile.
It is to note that the wavelength position of the \ktwov{} enhancement nearly remains the same in all the profiles throughout the evolution of the CBG.
The \lambdat diagram for the \CaIR line shows a small intensity enhancement at $t=75$~s in the blue wing of the \CaIR{} line.
The \CaIR line profiles at the peak of the CBG (see, panel~(e)) show a weak emission peak around $-256$~\mAA with line core redshifted ($\Delta\lambda=+$140~\mAA{}).
The evolution of the CBG is poorly captured in the \CaIR line compared to that in the \CaK{} line because the former has $\sim4.5$ times lower cadence. 
The \CaIR line may not have been recorded when the CBG was at its peak. Therefore, the weak intensity enhancement in the \CaIR line can be attributed to the lower cadence.

Figures~\ref{fig:shocksevolution2} and \ref{fig:shocksevolution3} show the evolution of the CBG in \roi{B} and C. 
The evolution of the CBG in \roi{B} and C is qualitatively very similar to that of in \roi{A}.
The blue wing images of the \CaK{} line show enhancement in intensity, and red wing (at $+153.3$~\mAA) images do not.
The \lambdat diagram of the \CaK and the \CaIR lines for the brightest pixel in CBG in panels (b) and (c) is similar to Fig.~\ref{fig:shocksevolution1}.
The intensity enhancement in \roi{B} near the blue wing ($-256$~\mAA) of the \CaIR line is significant, suggesting that observation of the \CaIR{} line coincided with the maximum phase of the CBG.
In contrast to the CBG in \roi{A}, the CBGs in \roi{B} and C show sub-structures.
The \blos{} map of \roi{B} and \roi{C} shows a photospheric field strength of about $-$27 and $-$64~G, respectively.
At $t=289$~s in \roi{B} and $t=165$~s in \roi{C}, the blue wing images of the \CaK line display multiple islands of intensity enhancements within the overall morphological structure of the CBGs.
An overview of the rest of the five \roi{}s is presented in appendix \ref{sect:supplementary}.

\subsection{Identification of the CBG like spectral profiles}
\label{subsect:classificationdataset}
\begin{figure}[htbp]
   \centering
   \includegraphics{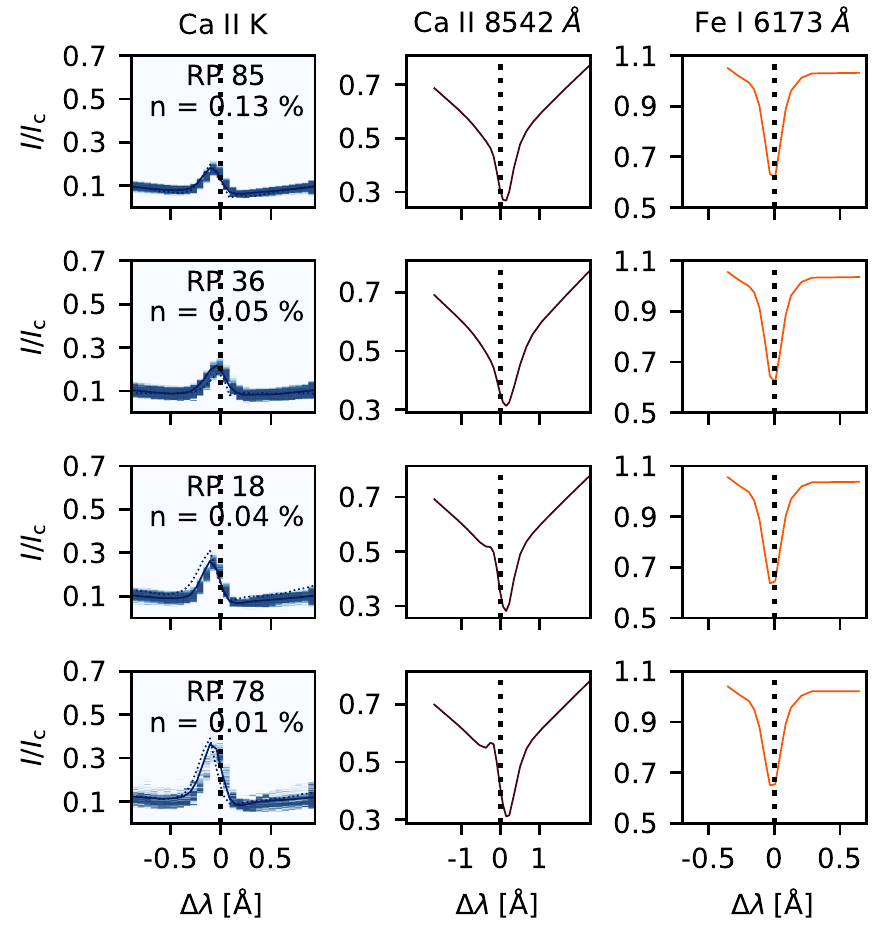}
      \caption{Illustrating the efficiency of the partitioning algorithm ($k$-means) in CBG clusters. The RPs in \CaK{}, \CaIR{} and \FeIline lines are shown using solid lines in blue, maroon and red colors, respectively. The dashed lines mark the zero position in the wavelength axis. The density distribution plots of the \CaK{} spectra for four RPs classified as CBG RPs are shown in blue color. The density (darker meaning higher concentration of spectra) corresponding to each \CaK RP shows the distribution of profiles over the entire time series. The percentage of profiles out of the whole dataset belonging to a particular cluster is denoted as $n$.
              }
         \label{fig:kmeansresult}
   \end{figure}

High resolution and high cadence spectropolarimetric solar observations typically generate large data volumes.
%
In the recent past, many authors have used algorithmic ways to reduce the dimensionality of the data, which facilitates efficient, qualitative and statistical analysis of millions of data points.
%
%
The $k$-means clustering is one such technique which has been extensively used to analyze solar spectra \citep[for eg.][]{2018ApJ...861...62P, 2019ApJ...875L..18S, 2021A&A...655A..28N, 2022arXiv220308172J}.
We employ the $k$-means technique to partition the \CaK{} spectral profiles into $k=100$ clusters as described in appendix \ref{sect:kmeansclustering}.
Each $k$-means cluster is represented by the mean of all the profiles belonging to that cluster which we refer to as the Representative Profile (RP).
The CRISP data have lower cadence and lower spatial resolution compared to CHROMIS, and therefore the evolution of the several CBGs could have been missed in the CRISP data as illustrated in Fig.~\ref{fig:shocksevolution1}.
Hence, we have performed the $k$-means clustering solely on the \CaK{} data from CHROMIS and not included the \CaIR{} and \FeIline{} lines from CRISP. 
We have projected the clusters over the co-spatial \CaIR{} and the \FeIline{} spectra to calculate corresponding RPs.
Out of the 100 RPs, we identified 4 RPs that show clear spectral signatures of CBGs in the \CaK{} line, a single peak emission around the nominal \ktwov{} and no emission about the \ktwor{} wavelength position (Sect. \ref{subsect:spatialspectral}).
We selected these 4 RPs based on the criteria that the intensity of the \ktwov{} emission peak must be double the intensity at the \konev{} wavelength position.
The selected 4 RPs in the \CaK{} line with the corresponding RPs in the \CaIR{} and the \FeIline{} are shown in Fig.~\ref{fig:kmeansresult}.
The RPs 78, 18, 36, and 85 from the \CaK{} data in decreasing order show the maximum enhancement in the \ktwov{} peak.
A hint of emission in the blue wing of the \CaIR{} line is seen only in RPs 78 and 18, whereas the RPs 85 and 36 are typical absorption profiles.
The corresponding \FeIline{} RPs do not show any peculiar spectral signatures.
The density distribution of the \CaK{} spectra within a particular cluster is shown by blue shaded areas in Fig.~\ref{fig:kmeansresult}, which is close to mean profiles in each cluster (shown in black), implying a good partitioning through the $k$-means algorithm.

\subsection{Inversions}
\label{subsect:inversionrps}

\begin{table*}[htbp]
\caption{Node positions in \logtau{} scale for temperature, line of sight velocity and microturbulence used for inversions of different group of RPs. } 
\label{table:inversionstrategy}
\begin{center}
\begin{tabular}{ c c c c  }
\hline\hline
Category & Temp & $V_{\mathrm{LOS}}$ & $V_{\mathrm{turb}}$\\
\hline
Quiescent & $-5$, $-4$, $-3$, $-2$, $-1$, 0 & $-5$, $-3$, $-1$ & $-5$, $-3.5$, $-1.5$, 0\\
Emission & $-5$, $-4$, $-3$, $-2$, $-1$, 0 & $-6$, $-5$, $-3.5$, $-1.5$, 0 & $-5$, $-3.5$, $-1.5$, 0\\
RP 18 / 78 & $-4.5$, $-3$, $-2$, $-1$, 0 & $-6$, $-5$, $-3.5$, $-1.5$, 0 & $-5$, $-3.5$, $-1.5$, 0\\
\hline
\end{tabular}
\end{center}
\end{table*}

\begin{figure}[htbp]
   \centering
   \textbf{\roi{A}}
   \includegraphics[]{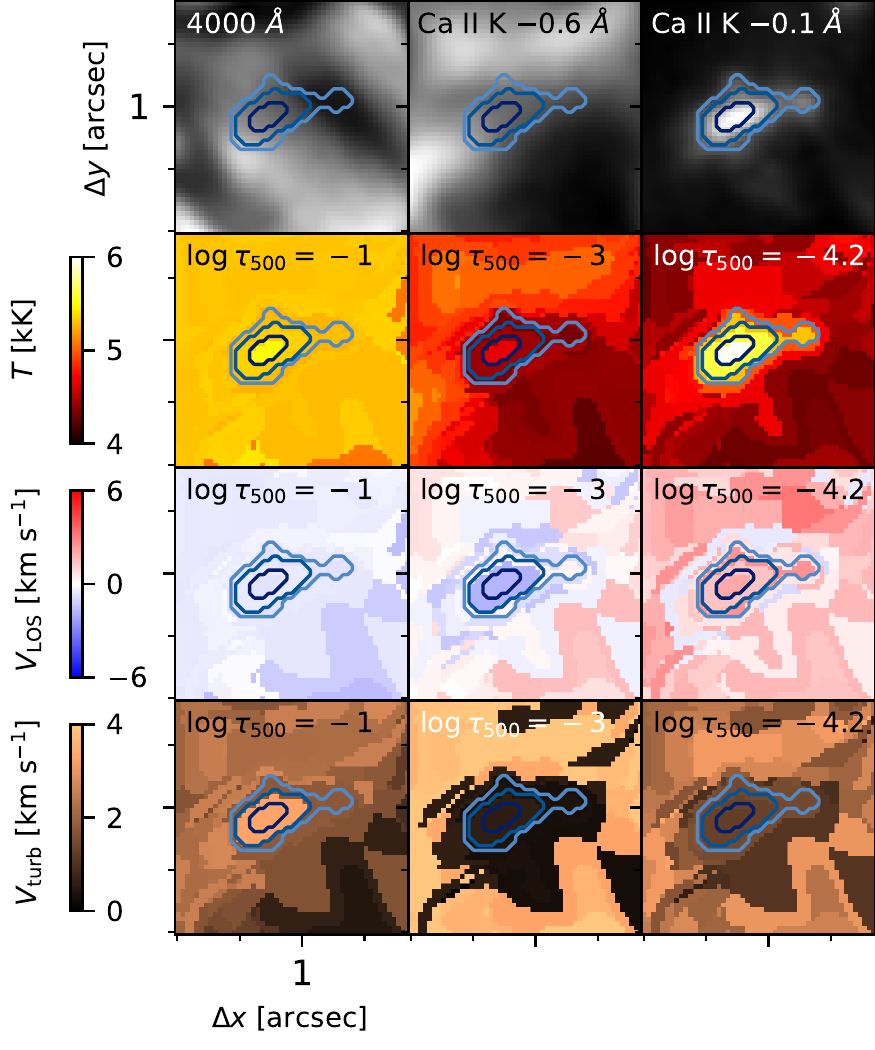}
      \caption{
       Maps of initial guess atmospheres for \roi{A} based on inversions of corresponding RPs. The first row shows images of $4000$~\AA\, continuum, \CaK{} line wing image ($-0.6$~\AA) and line core blue wing image ($-0.1$~\AA). The second, third and fourth rows show values at \logtau{} = $-1$, $-3$, and $-4.2$ of \temp{}, \vlos{} and \vturb{}, respectively. The outer contour (cyan) is made with pixels belonging to RPs 85, 36, 18, and 78, while the middle contour (azure blue) is made with RPs 18 and 78. The inner contour (navy blue) shows the region belonging to RP 78, the strongest CBG RP.
              }
         \label{fig:kmeansinitres}
   \end{figure}
We applied the MPI-parallel STockholm inversion Code \citep[STiC,][]{2016ApJ...830L..30D,2019A&A...623A..74D} to retrieve the atmospheric parameters and the evolution of the selected CBGs.
STiC is based on a modified version of the RH radiative transfer code \citep{2001ApJ...557..389U} and uses the cubic Bezier solvers to solve the polarized radiative transfer equation \citep{2013ApJ...764...33D}.
It can fit multiple spectral lines simultaneously in non-LTE assuming statistical equilibrium.
It takes into account partial re-distribution effects (PRD) using a fast approximation method \citep[for more details see][]{2012A&A...543A.109L}.
STiC fits the intensity in each pixel, assuming a plane-parallel atmosphere (also called the 1.5D approximation).
The equation-of-state utilized in STiC is obtained from the library functions in Spectroscopy Made Easy (SME) package code \citep{2017A&A...597A..16P}.

We have inverted the \CaK{}, \CaIR{} and \FeIline{} lines simultaneously.
We used a 6-level \ion{Ca}{ii} atom.
The \CaHK profiles are synthesized in PRD \citep{1974ApJ...192..769M, 1989A&A...213..360U}, and the \ion{Ca}{ii}~IR profiles are modeled using the complete re-distribution (CRD) approximation.
The atomic parameters for the \FeIline line are retrieved from Kurucz's line lists \citep{2011CaJPh..89..417K} and synthesized in the LTE approximation.
The line broadening cross-sections of radiative transitions of the \ion{Ca}{II} atom for collisions with neutral Hydrogen are retrieved from the table published by \cite{2000A&AS..142..467B}.
Hydrogen broadening Cross-Section Calculator (\mbox{abo-cross}\footnote{\url{https://github.com/barklem/abo-cross}}) code had been used to calculate line broadening cross-sections of the \FeIline{} line \citep{1995MNRAS.276..859A, 1997MNRAS.290..102B, 1998PASA...15..336B, 1998MNRAS.296.1057B}.

STiC iteratively perturbs the atmospheric parameters like temperature (\temp{}), line of sight velocity (\vlos{}), and microturbulence (\vturb{}) to minimize the $\chi^{2}$ distance between the synthesized and the observed profile.
The stratification of atmospheric parameters is given on the optical depth scale at 5000~\AA\, (500 nm) hereafter \logtau.
The gas pressure and density stratification is computed assuming hydrostatic equilibrium.
The atmosphere is perturbed at predefined specific locations of the \logtau\, scale, called nodes, then interpolated to the full depth grid.
Providing an initial guess atmospheric model close to the actual solution helps reduce the time it takes for inversion codes to converge.
Therefore, to find a good guess value of atmospheric parameters and finalize the node positions, we first invert the RPs.

We have only inverted the Stokes~$I$ of the \CaIR{}, and \FeIline{} along with the \CaK{} spectrum to infer the stratification of \temp{}, \vlos{} and \vturb{}.
We did not include the other Stokes parameters due to negligible signal and the selected \roi{s} having small magnetic activity.
Due to lower cadence of the CRISP data with respect to CHROMIS, we assigned one-fourth weightage to the \CaIR{} and one-half weightage to the \FeIline{} compared to the \CaK{} data from CHROMIS while inverting the RPs and the \roi{s}.
The quality of inversion fits are discussed in appendix \ref{sect:qualityoffits}.

To find suitable inversion node positions, we started with the FAL-C \citep{1985cdm..proc...67A, 1993ApJ...406..319F} model atmosphere as an initial guess for the inversions of the RPs.
We started with three equally spaced nodes in \temp{} and \vturb{} and one in \vlos{} and proceeded with inversions.
Step by step, we added nodes for those RPs for which we did not find a good fit with the initial setup.
After achieving a reasonable fit, we calculated response functions for all the RPs.
Response functions contain information of how a spectral profile is sensitive to perturbations in atmospheric parameters at different \logtau{}.
Depending on the response functions, we re-positioned nodes for different atmospheric parameters, and in case a satisfying fit was not achieved, we added more nodes.
With this practice, we concluded that different RPs require different number of nodes along with their different positioning along the optical depth scale for achieving the best fit.
Thus, we classified the RPs into three groups based on the number of nodes required and their position, one having ``quiescent'' profiles (30 RPs), one having ``emission'' profiles (68 RPs), and the third group consisting of 2 RPs, numbered 18 and 78, which are the strongest CBG RPs.
Quiescent RPs are those RPs which do not show any clear \ktwo{} features. 
Profiles that show either one or both the \ktwo{} features are referred to as emission RPs.
Table \ref{table:inversionstrategy} shows the node positions (in \logtau{} scale) used for different atmospheric parameters for the three categories of RPs.
%
We describe the guess atmospheric parameters inferred from the inversions of the RPs for one time step of the \roi{A} in Fig.~\ref{fig:kmeansinitres}. 
There is a temperature rise of about $1$~kK (at \logtau{} = $-$4.2) compared to neighboring pixels and a blueshift of about $-2$~\kms at \logtau{} = $-$3 at the pixels identified as CBG.
We used the guess atmosphere with the same node positions as the corresponding RPs to invert all time steps of all \roi{}s.
The inversions of the selected time sequences for all 8 \roi{}s (78 $\times$ 50 $\times$ 50 = 195,000 pixels) starting from the guess atmospheres derived above took 48750 CPU hours using a cluster with each node having two Intel\textsuperscript{\sffamily\textregistered} Xeon\textsuperscript{\sffamily\textregistered} E5-2683 v4 processors with 16 cores each and 2.10 GHz base frequency.

For the absolute velocity calibration, we refer to the convective blueshift of the \FeIline{} line at $\mu=0.7$ measured by \citet{2019A&A...624A..57L}.
We calculated a weighted average of the velocities inferred from the RPs in the \logtau range of [$-1$, $0$] with weights proportional to the number of occurrences of the RPs in the entire FOV and subtracted it from the convective blueshift to derive the calibration velocity.
Finally, we added the calibration velocity ($-1.12$ \kms) to the \vlos{} stratification inferred from the inversion of the \roi{}s.
  
\section{Results}
\label{sect:results}

\subsection{Evolution of atmospheric parameters}
\label{subsect:FOVanalysis}

\begin{figure*}[htbp]
\begin{center}
    \textbf{ROI A}
\end{center}
\begin{subfigure}{0.5\textwidth}
\resizebox{0.96\textwidth}{!}{\includegraphics[trim={0 {0.03\textwidth} 0 0},clip]{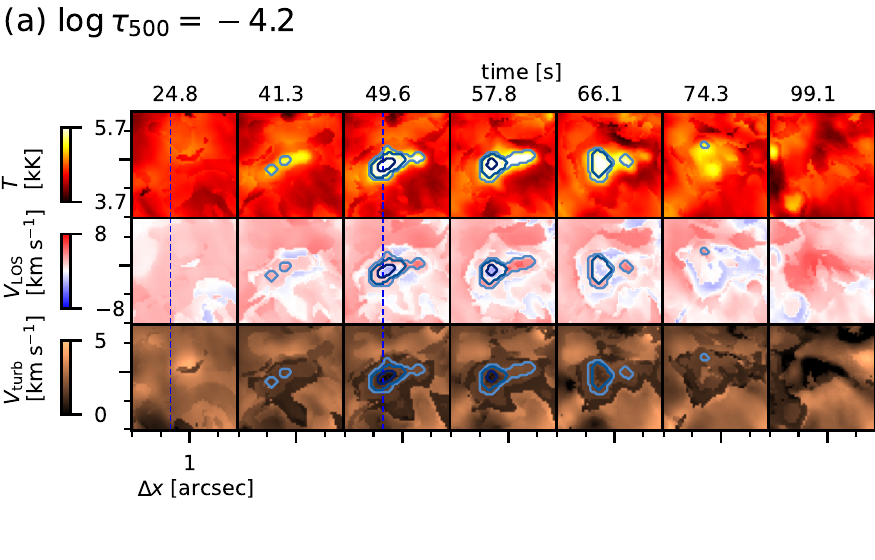}}
\resizebox{0.96\textwidth}{!}{\includegraphics[trim={0 {0.03\textwidth} 0 0},clip]{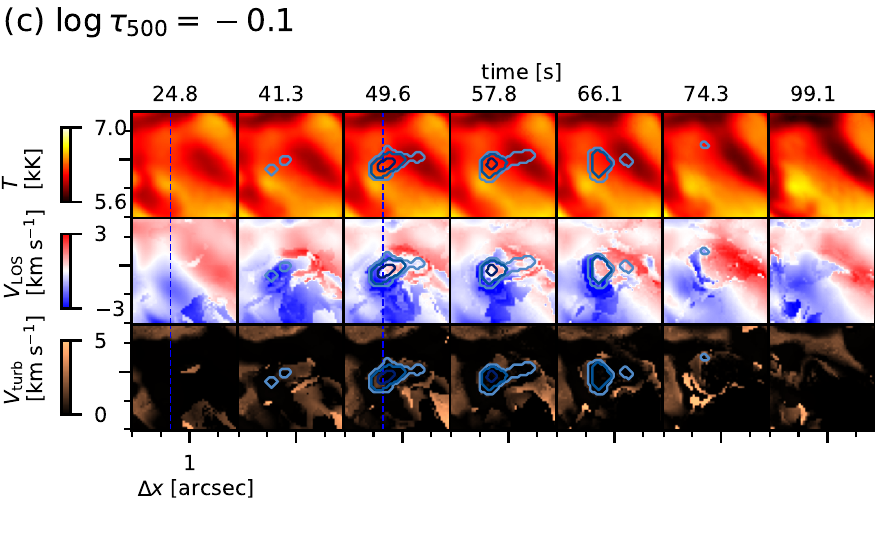}}
\end{subfigure}
\begin{subfigure}{0.5\textwidth}
\resizebox{0.96\textwidth}{!}{\includegraphics[trim={0 {0.03\textwidth} 0 0},clip]{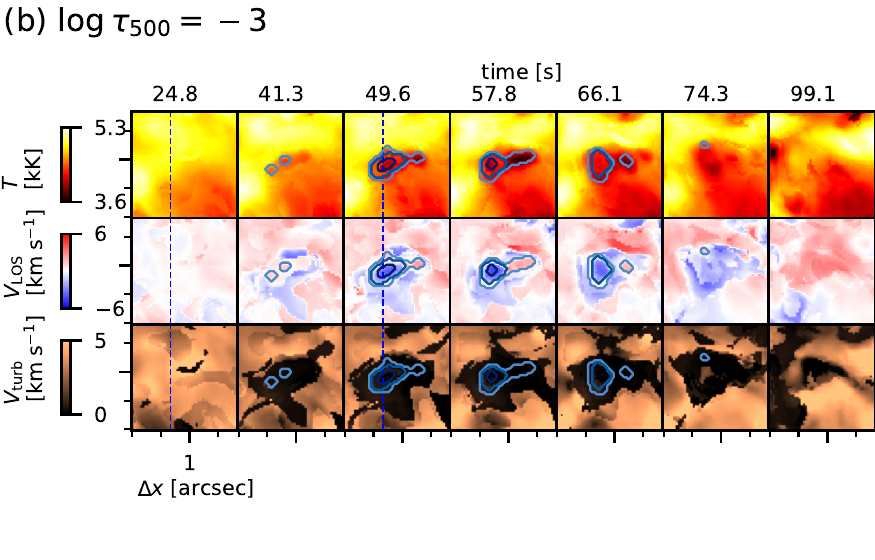}}
\resizebox{0.96\textwidth}{!}{\includegraphics[trim={0 {0.03\textwidth} 0 0},clip]{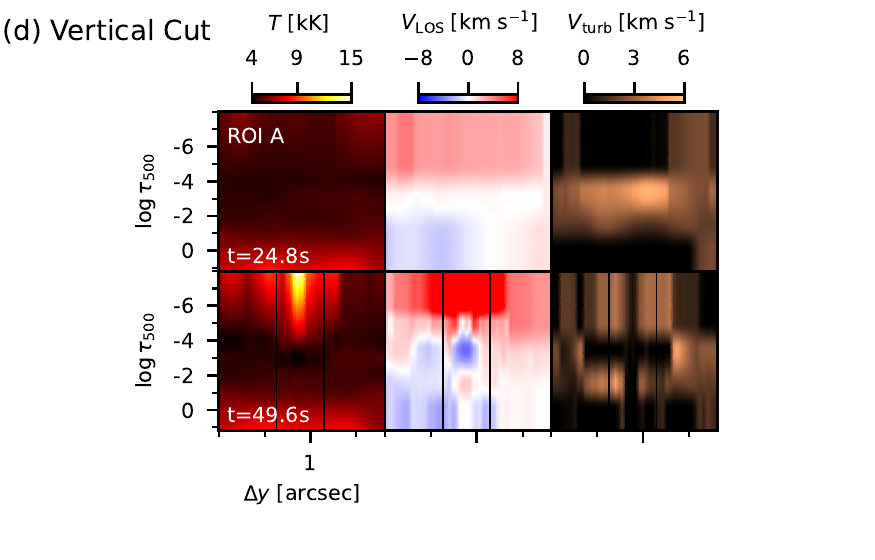}}
\end{subfigure}
\caption{Evolution of inferred stratified atmospheric parameters from the inversions of \roi{A}. The parameter maps for three values of \logtau = [$-4.2$, $-3$ and $-0.1$] are shown in the panels (a), (b) and (c), respectively. The first, second, and third rows show the evolution of \temp{}, \vlos{}, and \vturb{}. Panel (d) shows cross cuts along the depth of the atmospheres through the $y$-axis for two time steps as indicated. The $x$ positions for these cross cuts are indicated by blue dashed lines in panels (a)--(c). The contours in each panel show the pixels classified as CBGs as shown in Fig.~\ref{fig:shocksevolution1}.}
\label{fig:inversionmapa}
\end{figure*}

\begin{figure*}[htbp]
\begin{center}
    \textbf{ROI B}
\end{center}
\begin{subfigure}{0.5\textwidth}
\resizebox{0.96\textwidth}{!}{\includegraphics[trim={0 {0.03\textwidth} 0 0},clip]{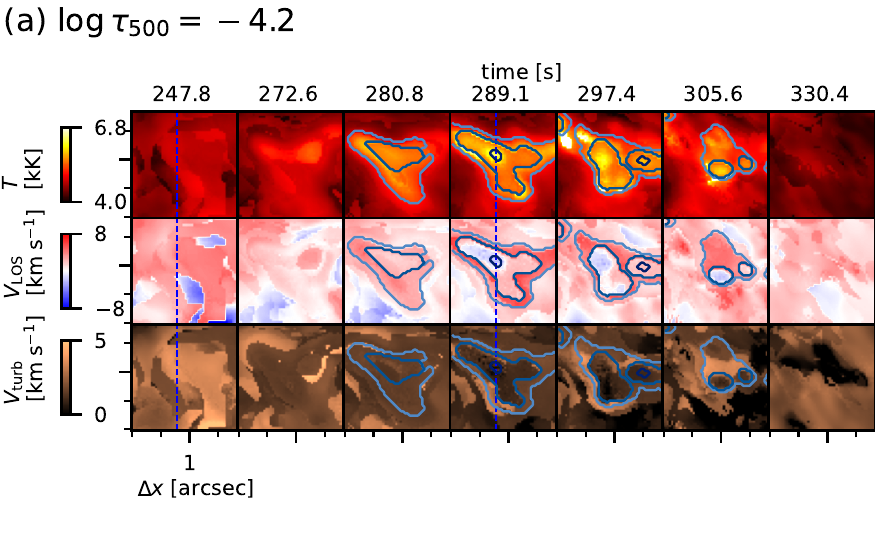}}
\resizebox{0.96\textwidth}{!}{\includegraphics[trim={0 {0.03\textwidth} 0 0},clip]{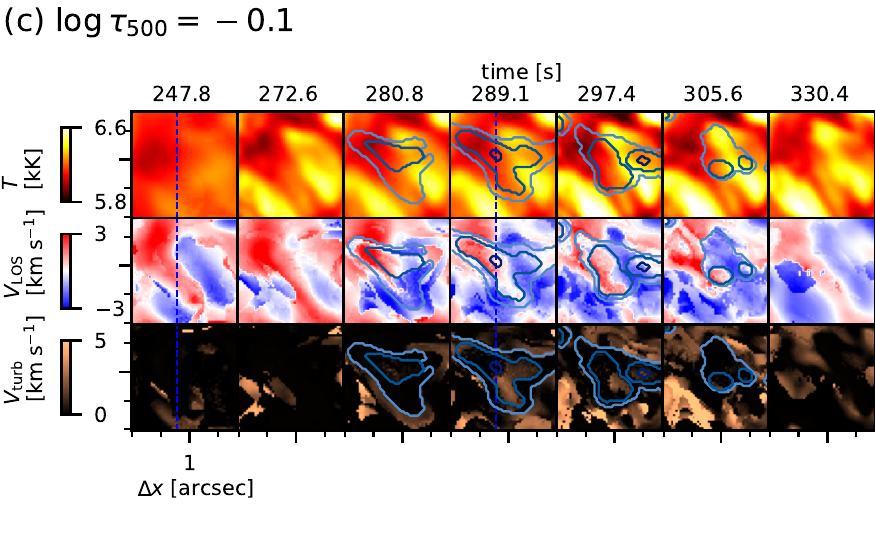}}
\end{subfigure}
\begin{subfigure}{0.5\textwidth}
\resizebox{0.96\textwidth}{!}{\includegraphics[trim={0 {0.03\textwidth} 0 0},clip]{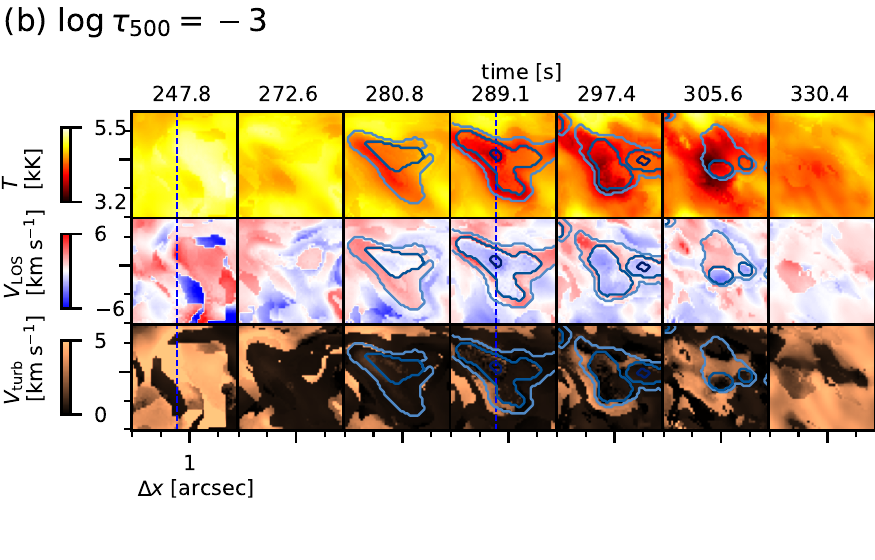}}
\resizebox{0.96\textwidth}{!}{\includegraphics[trim={0 {0.03\textwidth} 0 0},clip]{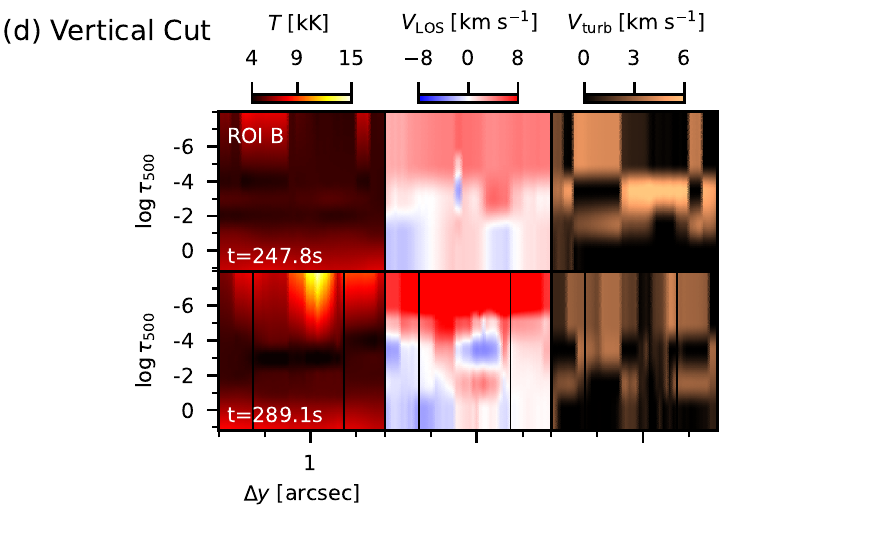}}
\end{subfigure}
\caption{same as Fig.~\ref{fig:inversionmapa} for \roi{B}.}
\label{fig:inversionmapb}
\end{figure*}

\begin{figure*}[htbp]
\begin{center}
    \textbf{ROI C}
\end{center}
\begin{subfigure}{0.5\textwidth}
\resizebox{0.96\textwidth}{!}{\includegraphics[trim={0 {0.03\textwidth} 0 0},clip]{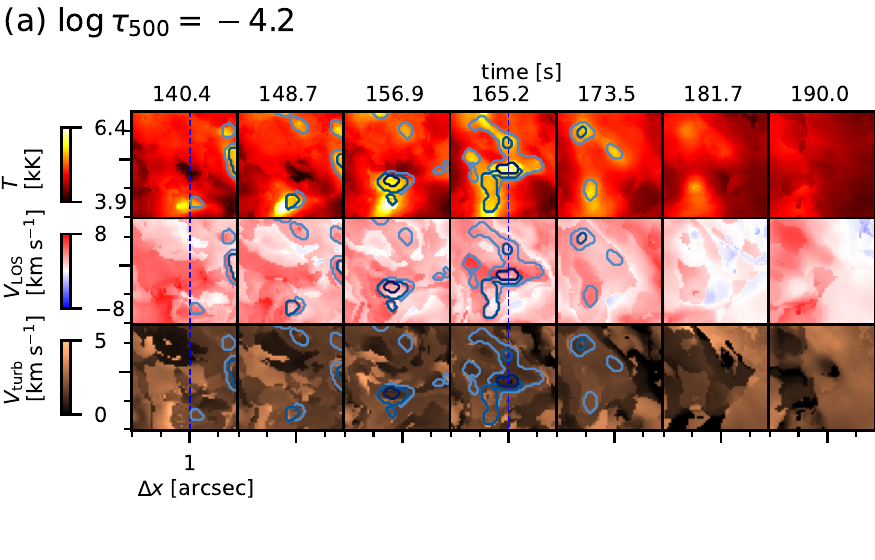}}
\resizebox{0.96\textwidth}{!}{\includegraphics[trim={0 {0.03\textwidth} 0 0},clip]{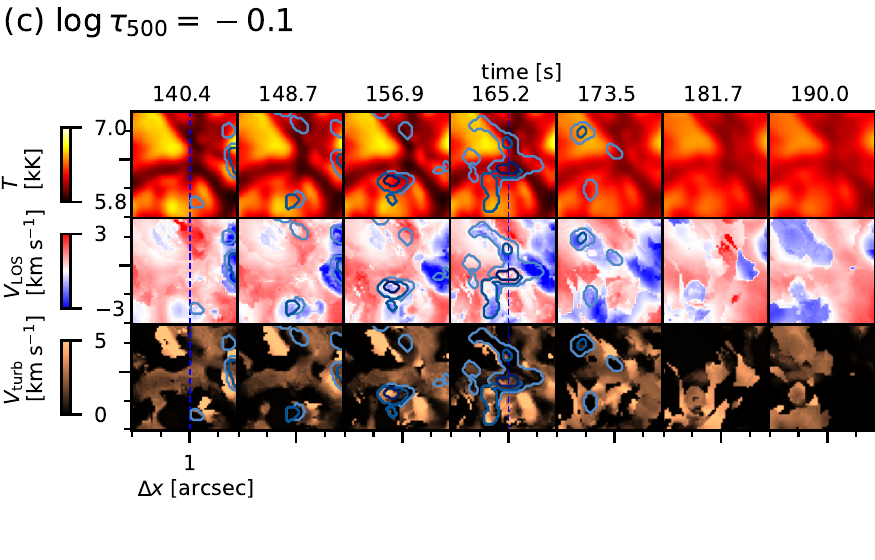}}
\end{subfigure}
\begin{subfigure}{0.5\textwidth}
\resizebox{0.96\textwidth}{!}{\includegraphics[trim={0 {0.03\textwidth} 0 0},clip]{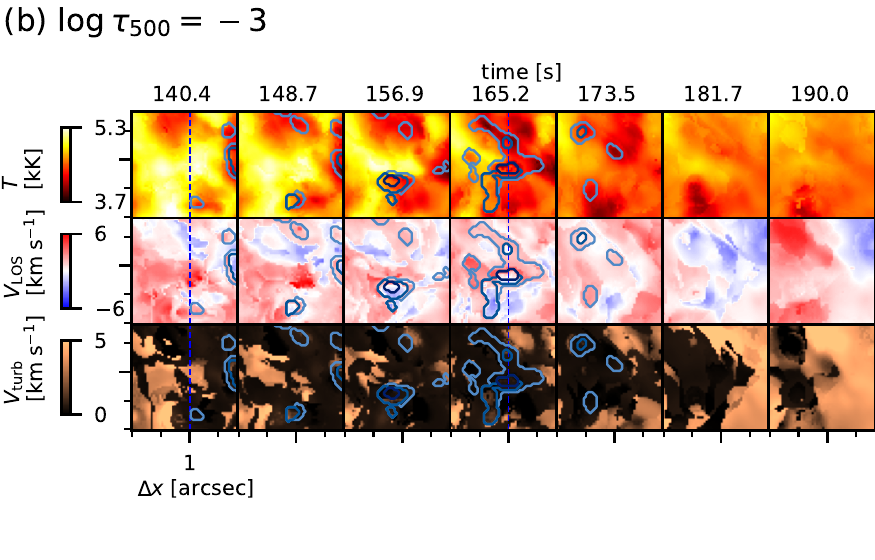}}
\resizebox{0.96\textwidth}{!}{\includegraphics[trim={0 {0.03\textwidth} 0 0},clip]{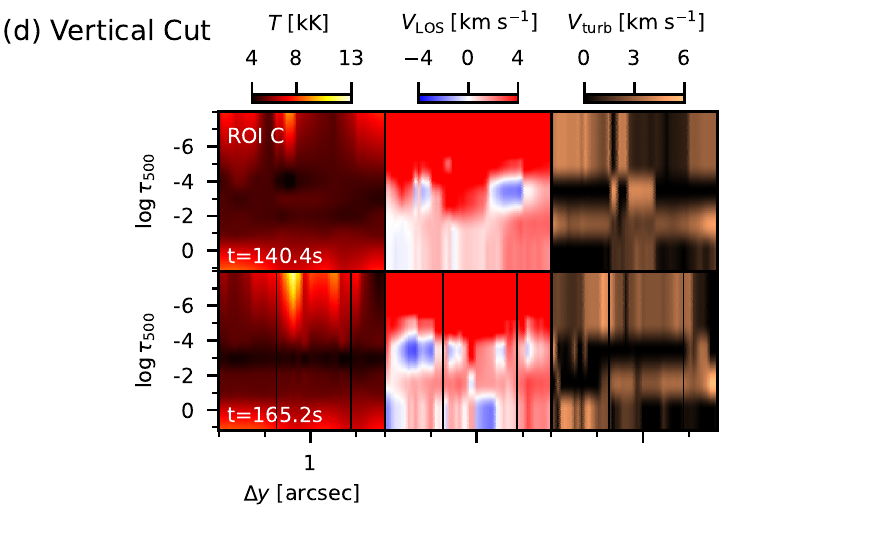}}
\end{subfigure}
\caption{same as Fig.~\ref{fig:inversionmapa} for \roi{C}.}
\label{fig:inversionmapc}
\end{figure*}

We present the evolution of the atmospheric parameters inferred with the inversions of the eight ROIs that are marked in Fig.~\ref{fig:FOV}.

We should note that the data presented above was recorded at a viewing angle of $\mu$ = 0.7 with solar normal.
The stratification of the atmosphere inferred from inversions may be affected by this geometrical effect because the spatial resolution of the data ($0\farcs038\sim25$~km) is extreme compared to the typical formation height of the shocks ($\sim$1000~km) \citep{1997ApJ...481..500C}.
If one assumes that the shocks propagate vertically upwards \citep{1997ApJ...481..500C, 2009A&A...494..269V}, non-shocking atmosphere may have some contribution to the inferred stratification of atmospheric parameters of the CBGs when viewed from a slanted geometry.

The inversion results for \roi{A} are displayed in Fig.~\ref{fig:inversionmapa}. 
Before the onset of the CBG, the retrieved \temp{} map at \logtau{} = $-$4.2 looks homogeneous with minimal variations in \temp{} across the ROI and with an average value of $\sim3.8$~kK.
At the onset of the CBG, at $t=41.3$~s, a weak compact enhancement in $T$
appears at the center of the ROI.
The CBG reaches the maximum area at $t=49.6$~s where clear structures of enhancements in \temp{} can be seen.
At the peak of the CBG, these enhancements in \temp{} are of about $1$--$2$~kK, and these enhancements are found maximum at the core of the CBG (inner contour in navy blue color).
After $t=66.1$~s the area and \temp{} enhancements start decreasing.
In contrast, the \temp{} map at \logtau{} = $-$3 shows a reduction in \temp{} at the location of CBG.
At the peak of the CBG ($t=49.6$~s), this decrease in \temp{} is maximum in the range of about $0.5$--$1$~kK.

At \logtau{} = $-4.2$, before the onset of the CBG, the \roi{} in general shows weak downflow.
Between $t=49.6$--66.1~s, we see upflows in the CBG at the locations corresponding to higher enhancements in \temp{}.
The retrieved \vlos{} exhibits stronger upflows in the CBG at \logtau{} = $-3$ compared to that of at \logtau{} = $-4.2$.
At \logtau{} = $-4.2$ and $-3$, the maximum upflow is about $-4$~\kms{} and $-6$~\kms{}, respectively.
In the photospheric layer, at \logtau = $-$0.1, a typical granulation pattern is visible in the \temp{} maps.

At \logtau{} = $-4.2$, the pixels corresponding to the CBG, in general, show less microturbulence compared to surrounding pixels.
Moreover, at \logtau{} = $-3$ and $-$0.1, there is almost no microturbulence at the CBG.
In contrast, we see constant and consistent microturbulence of $3$--$5$~\kms{} outside of the CBG.
One could argue that the \temp{} increase inferred in the CBGs atmosphere is an artifact caused by such low values of microturbulence.
Inversions are known to have degeneracy between \temp{} and \vturb{} as competing mechanisms for increasing the width of the spectral line.
\citet{2022A&A...659A.165D} demonstrated that when the line is in emission, the \temp{} and the \vturb{} are anti-correlated, which means that increasing the \temp{} broadens the spectral line, so to maintain the same width the \vturb{} must be decreased.
We note here that the enhancement in \temp{} at \logtau{}$\simeq-$4 is necessary to produce intensity enhancement at the \ktwov{} wavelength position; only uncertainty in the amplitude of the enhancement in the \temp{} due to such low values of \vturb{} is not very well determined.
To better understand the degeneracy between \temp{} and \vturb{}, we performed an experiment in which we inverted the brightest pixel of \roi{B}, but with different values of the \vturb{}, as described in appendix~\ref{sect:tvturbrel}.
We concluded that as the value of the \vturb{} increases, the intensity at the \ktwov{} wavelength position decreases, causing the fit of the \CaK{} profile to worsen.
For higher values of \vturb{}, the inverted profiles show the \ktwor{} peak which is absent in the observed profile.
Thus, minimal values of \vturb{} are required to achieve a satisfactory fit of the \CaK{} profile, and the uncertainty in \temp{} enhancement due to the degeneracy between \temp{} and \vturb{} is not very significant.

The vertical cut before the onset and at the peak of the CBG activity is shown in panel~(d).
The \temp{}, compared to the pixels outside CBG, starts to rise during the CBGs at \logtau $\simeq-$3.8 which is colocated with the maximum upflows in \vlos{}.
These upflows are of about $-4$ \kms with a temperature increase of about $1$ kK with respect to before the onset of the CBG.
The atmosphere before the occurrence of the CBG displays a weak downflow at \logtau = $-4$ and in the atmosphere above.
During the CBGs, the magnitude of these downflows show enhancements simultaneous with the enhancements in \temp{} and the appearance of upflows between \logtau = $-3$ and $-4$.
At the core of the CBG, no microturbulence is present at any continuum optical depth.

The stratification and evolution of atmospheric parameters of the CBG in \roi{B} shown in Fig.~\ref{fig:inversionmapb} is similar to \roi{A}.
Before the onset of the CBG, the atmospheres at and above \logtau = $-4.2$ show a weak downflow.
The start of the enhancement in \temp{} relative to surrounding pixels is colocated with upflows at \logtau $\simeq-3.8$ (see panel~(d) of Fig.~\ref{fig:inversionmapb}).
There is relatively less microturbulence in the CBG than the surrounding pixels at all optical depths.

The structure and evolution of the atmosphere obtained of the CBG in \roi{C} is shown in Fig.~\ref{fig:inversionmapc}.
Contrary to ROIs A and B, the inferred \vlos{} maps at the location of enhancements in \temp{} show no signature of upflows at \logtau{} = $-4.2$.
The retrieved \temp{} map of the CBG at \logtau{} = $-$4.2 shows sub-structures displaying localized $T$ enhancement.
These substructures do not have a consistent structure in the \vlos{} map at \logtau{} = $-3$, some of them show upflow and others downflow.
In contrast to the stratification of the \vlos{} shown for \roi{A} and B, we see downflows between \logtau{} = $-3.5$ and $-4$ at the location of maximum \temp{} enhancement in the CBG (see panel~(d)).

The structure and evolution of retrieved atmospheric parameters of the rest of the 5 ROIs are presented in the appendix \ref{sect:supplementary}.

\begin{figure*}[htbp]
\begin{subfigure}{\textwidth}
\begin{center}
\resizebox{0.9\textwidth}{!}{\includegraphics[]{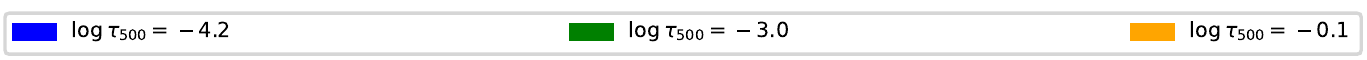}}    
\end{center}
\end{subfigure}
\begin{subfigure}{\textwidth}
\resizebox{0.24\textwidth}{!}{\includegraphics[]{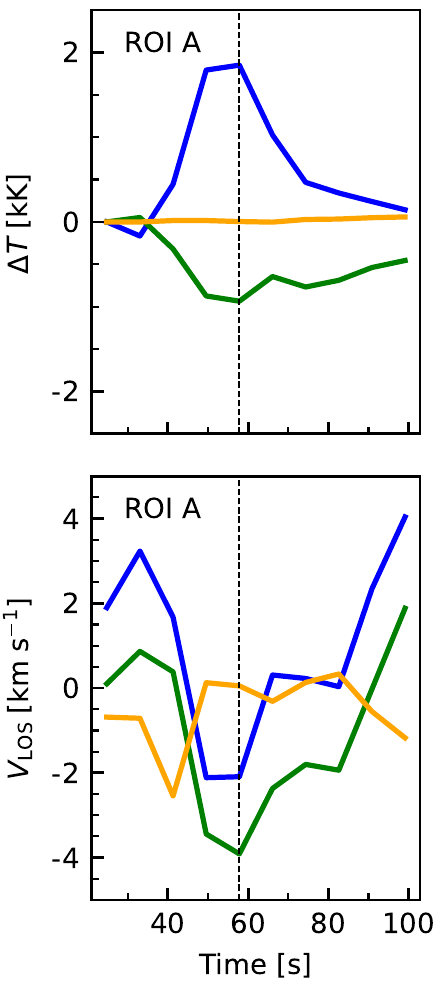}}
\resizebox{0.24\textwidth}{!}{\includegraphics[]{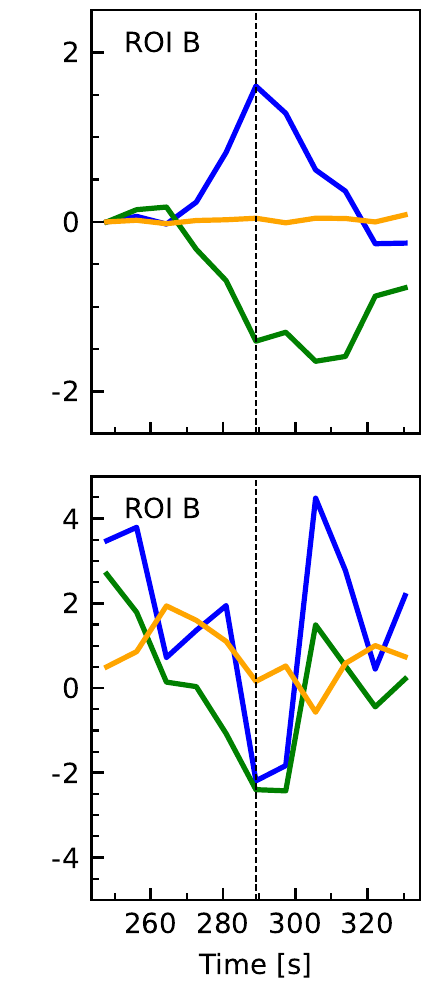}}
\resizebox{0.24\textwidth}{!}{\includegraphics[]{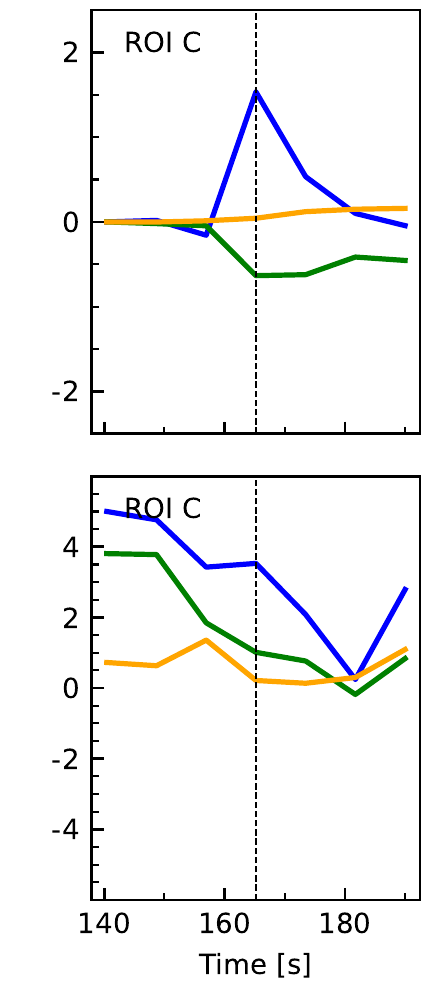}}
\resizebox{0.24\textwidth}{!}{\includegraphics[]{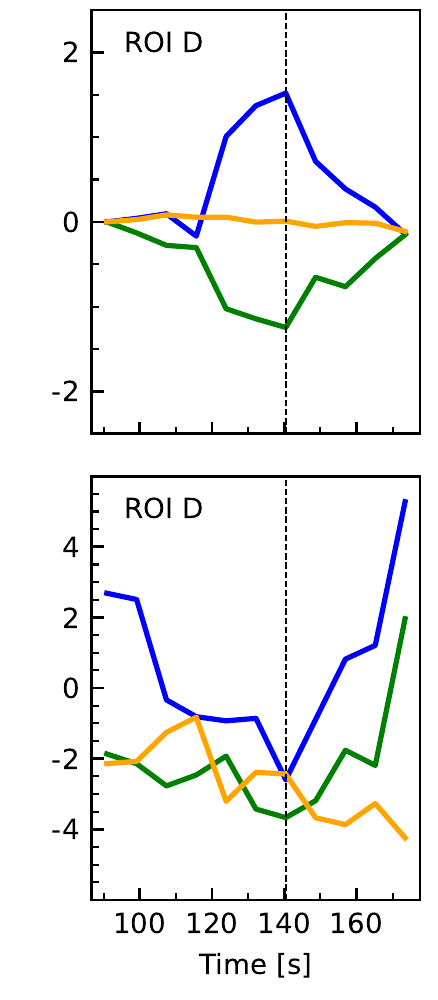}}
\end{subfigure}
\begin{subfigure}{\textwidth}
\resizebox{0.24\textwidth}{!}{\includegraphics[]{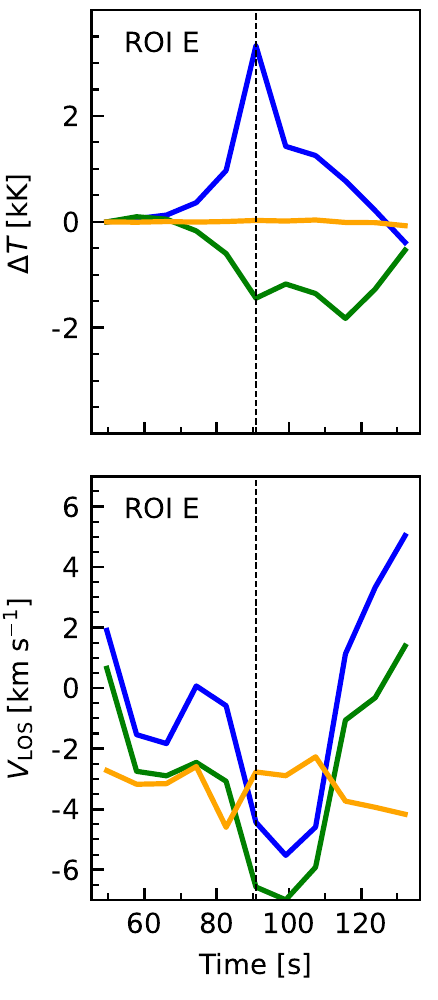}}
\resizebox{0.24\textwidth}{!}{\includegraphics[]{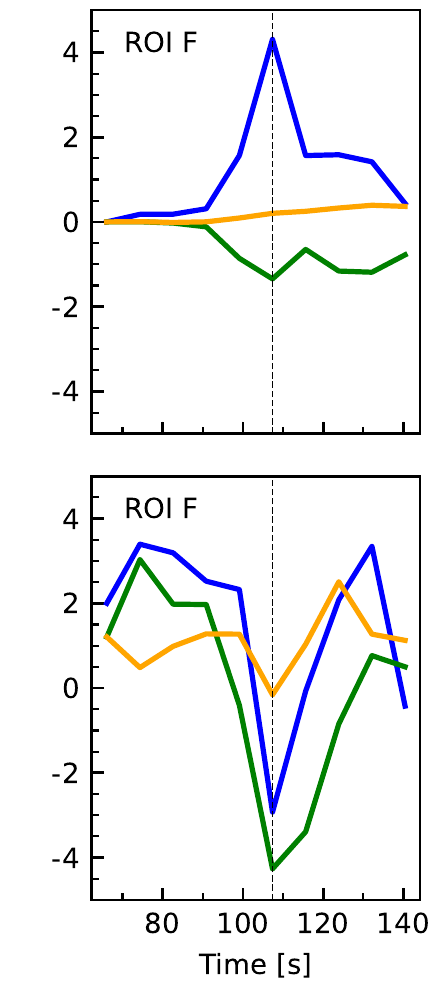}}
\resizebox{0.24\textwidth}{!}{\includegraphics[]{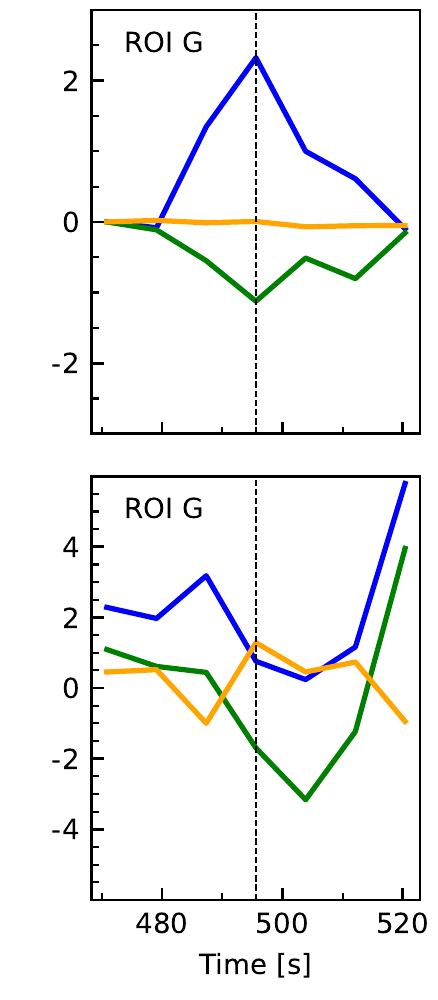}}
\resizebox{0.24\textwidth}{!}{\includegraphics[]{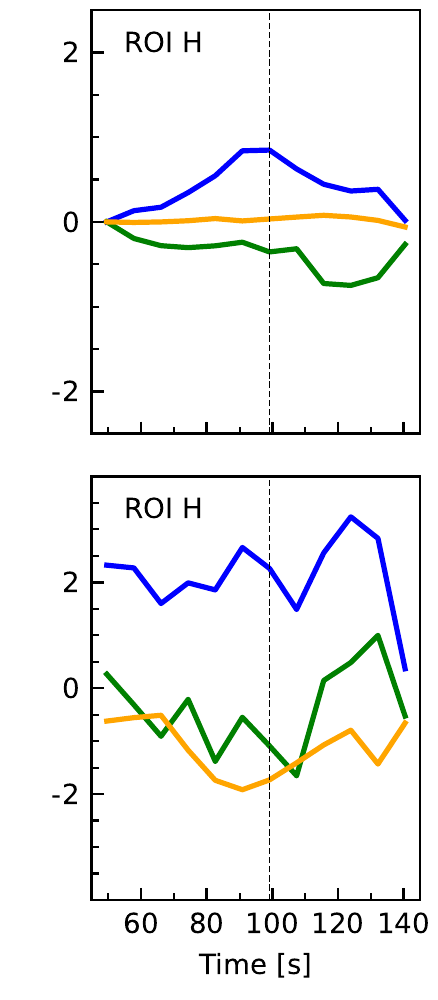}}
\end{subfigure}
    \caption{Time variation of \temp{} and \vlos{} of a brightest pixel of the CBGs in ROI A--H. For example the brightest pixel is marked by '+' in panel (a) of Fig.~\ref{fig:shocksevolution1}. The evolution of atmospheric parameters are shown at \logtau{} = $-$4.2, $-3$ and $-$0.1. The vertical line indicate the time of maximum \temp{} enhancement.}
    \label{fig:timevariation1}
\end{figure*}

In Fig.~\ref{fig:timevariation1}, we present the variation of the \temp{} and \vlos{} with time in the core of the CBG (also the brightest pixel) each from \roi{A--H}.
For example the brightest pixel is marked with '+' in panel (a) of  Fig.~\ref{fig:shocksevolution1}.
During the evolution of a CBG, the \temp{} at \logtau{} = $-4.2$ increases and attains a maximum enhancement with respect to the atmosphere before the onset of the CBG before falling to the values before the onset.
This enhancement in \temp{} in general is about $1.5$--$2$~kK.
The inferred \vlos{} at \logtau{} = $-4.2$ at the time of large temperature enhancement in \temp{} in the core of the CBGs show the signature of upflows.
For example, the brightest pixel in \roi{E} and F shows maximum enhancement in \temp{} of about $3.5$ and $4.5$~kK with upflows of about $-5.5$ and $-3$~\kms, respectively.
However, sometimes these signature of upflows are missing.
For example, in \roi{G} and H, the maximum enhancement in \temp{} is of 2.5 and 1~kK with downflows of about $+$0.5 and $+$2~\kms{}, respectively.
The evolution of \temp{} at \logtau{} = $-3$ is opposite to that of at \logtau{} = $-4.2$, as the CBG evolves, the \temp{} decreases, attains minimum, and then recovers to the approximate value before the onset.
The retrieved \vlos{} at \logtau{} = $-3$ majorly show signature of upflows, however, sometimes of weak downflows (\roi{C}).
The upflows at \logtau{} = $-3$ at the time of peak enhancement in \temp{} are stronger than that of at \logtau{} = $-4.2$.

\subsection{\CaK{} response functions}
\label{subsect:cakresponsefunction}

\begin{figure*}[htbp]
   \sidecaption
   \includegraphics[width=12cm]{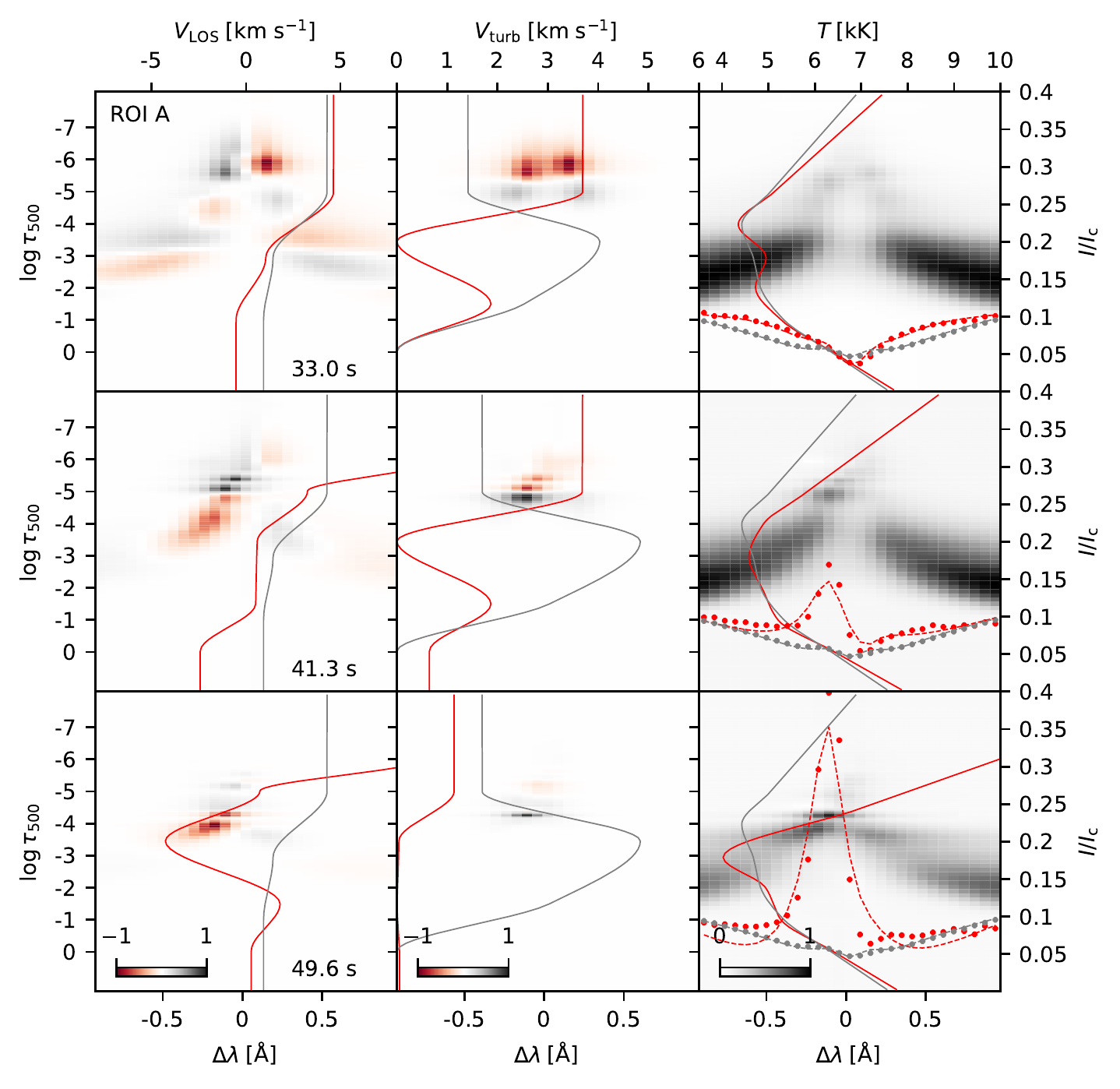}
      \caption{
      Normalised Response Functions of the \CaK line to the perturbations of $V_{\mathrm{LOS}}$ (the left column), $V_{\mathrm{turb}}$ (the middle column) and $T$ (the right column) for the marked pixel in Fig.~\ref{fig:shocksevolution1}(a) at three time steps namely,  $t=33.0$~s (quiescent phase, the first row), $t=41.3$~s (intermediate CBG phase, the second row) and $t=49.6$~s (when the CBG is at its maximum brightness, the third row); The inferred atmospheric quantities are shown with a red solid curve and the median profile over the entire FOV is shown in grey color for comparison. The observed intensities are shown with red (CBGs) and grey (median profile) filled circles in the right column and the fitted line profiles are shown with dashed curves. A gamma correction with $\gamma$=0.5 has been applied on the \temp{} response function before display.
              }
         \label{fig:responsefunc}
   \end{figure*}

To study the optical depths of formation of the CBGs, we present the response functions of the \CaK line with \vlos{}, \vturb{} and \temp{} in Fig.~\ref{fig:responsefunc}, in the same format as shown in \cite{2020A&A...637A...1K}.
\cite{1975SoPh...43..289B} define the response function (RF) for a physical parameter $X$ as $RF_{X}(\tau, \lambda)\;=\;\delta I(\lambda)/\delta X(\tau)$.

In the following part of this section, we discuss the response of the \CaK{} Stokes~$I$ profile to perturbations in $T$, \vlos{} and, \vturb{} in detail.
We have presented response functions for three time steps (row-wise) for the brightest pixel of the CBG in \roi{A}, i.e. before the onset of the CBG, at the onset, and when the CBG is at its maximum.
Each plot in 3$\times$3 grid shows the response of the Stokes~$I$ profile to perturbation in an atmospheric parameter (\vlos{}, \vturb{} and \temp{} column-wise) as a function of \logtau{} and wavelength.
The inferred atmospheric parameters from the inversions of the observed profile (red color) and median profile (gray color) are also shown for comparison.
The observed Stokes~$I$ profile (dotted red) and the median profile (dotted gray) are shown in the right column.
The synthesized intensities derived from inversions for observed and median profile are shown in red and gray dashed curves, respectively.

$RF_{T}$: At $t=33.0$~s, before the onset of the CBG, the stratification of $T$ retrieved from the multi-line inversions is similar to the \temp{} stratification obtained from the median profile.
The \CaK line within the observed spectral range is sensitive to perturbations in \temp{} at \logtau{} between $-6$ and $-2$ with line wings forming between \logtau{} = [$-4$, $-2$] and line core forming between \logtau{} = [$-6$, $-4$].
As the CBG evolves, at $t=41.3$~s, the \ktwov{} peak has a strong response around \logtau{} = $-5$ where we see an enhancement of \temp{} of $\sim0.8$~kK compared to that of in the atmosphere before the onset of the CBG.
The optical depth where the response to \temp{} is maximum has decreased from \logtau{} = $-5$ to \logtau{} = $-4.2$ at the time when we see the strongest intensity enhancement of the \ktwov{} peak (at $t=49.6$~s) that corresponds to $\sim2$~kK rise in \temp{}.
The response to \temp{} between \logtau{} = [$-4.2$, $-3$] has an asymmetric behavior about the \CaK{} line core with a higher response towards the blue wing.

$RF_{V_{\mathrm{LOS}}}$: Before the onset of the CBG (at $t=33.0$~s), the \CaK{} line is sensitive to a perturbation in \vlos{} in wide range of \logtau{} values between $-7$ and $-2$.
As the CBG evolves, at $t=41.3$~s when the \ktwov{} peak is visible, the maximum response to perturbations in \vlos{} is restricted between \logtau{} = [$-5.5$, $-3.5$] and shifted to the nominal \ktwov{} position.
There is weak response to perturbations in \vlos{} in the upper layers (between \logtau{} = [$-7$, $-5.5$]).
The maximum response to perturbations in \vlos{} further gets restricted between \logtau{} = [$-4.5$, $-3.5$] at the time (at $t=49.6$~s) of strongest enhancement of the \ktwov{} peak intensity.
The inferred \vlos{} near the optical depth with strongest response (\logtau{} $\simeq-3.9$) shows an upflow of $\sim-3$~\kms{}.

$RF_{V_{\mathrm{turb}}}$: Before the onset of the CBG (at $t=33.0$~s) the line is sensitive to the perturbations in \vturb{} between \logtau{} = [$-6.5$, $-4.5$] with the inferred \vturb{} of $\sim3.8$~\kms, that is higher than that from median profile ($\sim1.3$~\kms).
During the onset of the CBG (at $t=41.3$~s), the response to perturbations in \vturb{} gets limited between \logtau{} = [$-5.5$, $-4$] with inferred \vturb{} remains nearly unchanged.
When the enhancement in intensity of the \ktwov{} peak is the strongest (at $t=49.6$~s), the maximum response to perturbations in \vturb{} is restricted about \logtau{} = $-4$ and inferred \vturb{} reduces to $\sim1$~\kms above \logtau{} = $-4$ and vanishes below.

In general, the \CaK{} line is sensitive to perturbations in \temp{} between \logtau{} = [$-6$, $-2$], however, during the strongest enhancement in the \ktwov{} peak intensity, the sensitivity to perturbations gets restricted to atmosphere below \logtau{} = $-4.2$, which is also the optical depth of the start of the rise in \temp{}.
The sensitivity to perturbations in \vlos{} gets restricted between \logtau{} = [$-4.5$, $-3.5$] during the strongest enhancement of the \ktwov{} peak intensity, while before the onset of the CBG with no \ktwo{} features the \vlos{} is sensitive to wide range of optical depths between \logtau{} = [$-7$, $-2$].
The maximum response to perturbations in \vturb{} during the maximum enhancement of the \ktwov{} peak intensity is limited about \logtau{} = $-4.2$, with weak sensitivity up to \logtau{} = $-$5.5, while before the onset of the CBG, the sensitivity to perturbations in \vturb{} is limited between narrow range of \logtau{} = [$-6.5$, $-4.5$].
Although we have shown here the response of the \CaK{} line to the perturbations of \temp{}, \vlos{} and \vturb{}, we have included the \FeIline{} and \CaIR{} lines in inversions, which gives us confidence in the stratification of inferred atmospheric parameters from the photospheric layers (\logtau{} $\simeq0$) to the chromosphere (\logtau{} $\simeq-5$).

\subsection{Relationship of velocity and temperature enhancement}
\label{subsect:velocitytemprelationship}

\begin{figure*}[htbp]
    \centering
         \resizebox{0.86\textwidth}{!}{\includegraphics[]{images/legends_6.pdf}}
          \resizebox{0.48\textwidth}{!}{\includegraphics[]{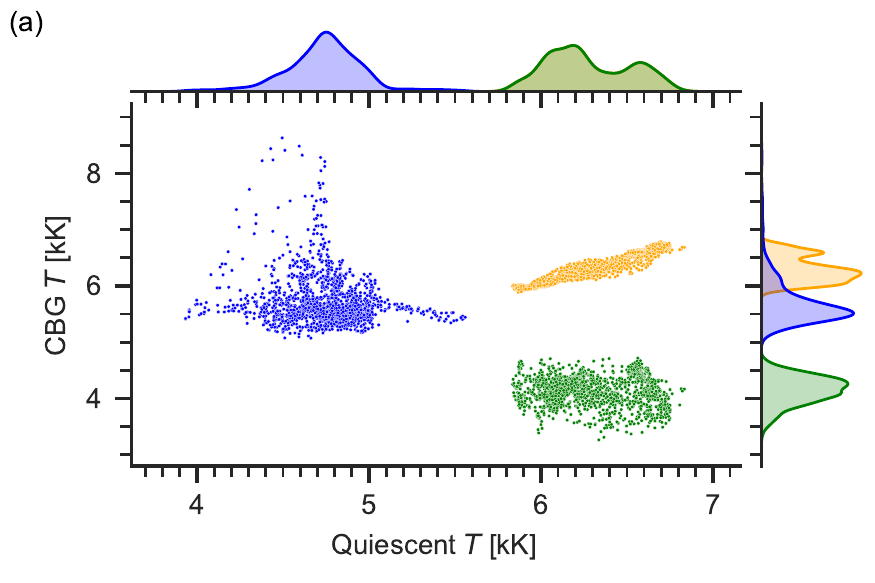}}
          \resizebox{0.48\textwidth}{!}{\includegraphics[]{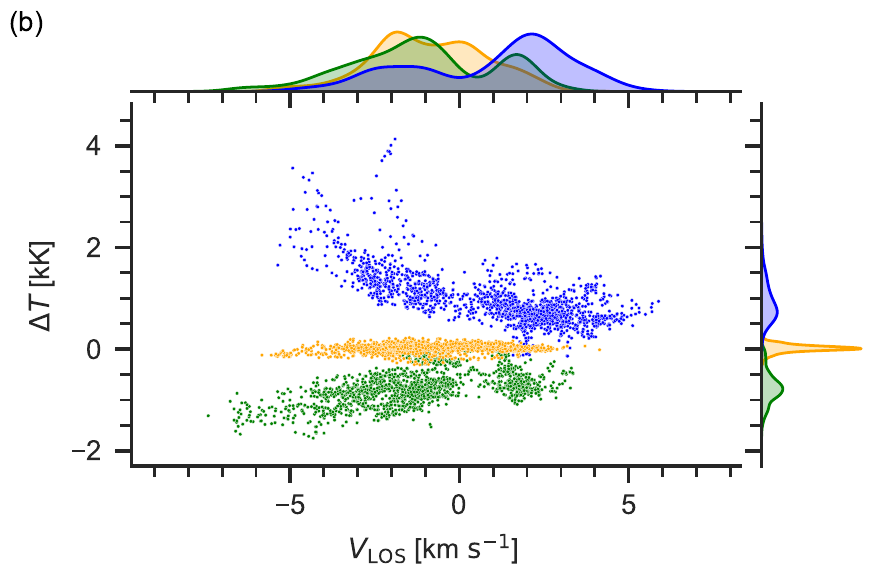}}
    \caption{Relationship of \temp{} enhancement and \vlos{} at \logtau{} = [$-$4.2, $-$3, $-$0.1]: Panel(a) shows the scatter plot between the \temp{} when the CBG is at its maximum brightness (maximum enhancement in the \ktwov{} peak intensity) with respect to the \temp{} before the onset of the CBG and panel (b) shows the scatter plot between the \temp{} enhancement with respect to the \vlos{} when the CBG is at its maximum brightness.}
    \label{fig:preshockpeakshock}
\end{figure*}

We present the scatter and probability distribution plots for maximum \temp{} in the CBGs with respect to the \temp{} before the onset of the CBGs in panel (a) of Fig.~\ref{fig:preshockpeakshock}.
Panel (b) of Fig.~\ref{fig:preshockpeakshock} presents the scatter and probability distribution plots for maximum \temp{} enhancement ($\Delta T$) in the CBGs with respect to the inferred \vlos{} at the time of the maximum enhancement of \temp{}.
We used the pixels belonging to RPs 18/78 in \roi{A--H}, for example pixels shown in middle contours (azure blue) in panel (a) of Fig.~\ref{fig:shocksevolution1}, to make the above plots.

At \logtau{} = $-$4.2, most pixels classified as CBGs have a near-constant maximum \temp{} of $\sim5.5$~kK irrespective of the \temp{} before the onset of the CBG.
Few pixels show a large \temp{} enhancement with maximum \temp{} values reaching up to $8.5$~kK.
The average quiescent \temp{} and enhancement in \temp{} is about $4.75$~kK and $0.75$~kK, respectively.
The \temp{} at \logtau{} = $-3$ for pixels in CBG is typically less than the \temp{} before the onset of the CBG with average decrease of $-0.75$~kK and a maximum reduction in \temp{} of $-$1.5~kK.
The \temp{} at \logtau{} = $-$0.1 nearly remains identical to that of before the onset of CBGs.

The \vlos retrieved at \logtau{} = $-4.2$ in general show upflows for $\Delta T$ values greater than 1~kK and downflows when $\Delta T$ is between 0.5--1~kK.
%
Stronger the upflows, higher the value of $\Delta T$.
The inferred \vlos{} at \logtau{} = $-3$ show upflows typically, with stronger upflows corresponding to more negative values of $\Delta T$.
At the photospheric layers, at \logtau{} = $-$0.1, the $\Delta T$ nearly remains zero irrespective of the value of \vlos{}.

\section{Discussions and conclusions}
\label{sect:conclusions}

We presented the evolution of stratified atmospheric parameters, i.e.,  temperature, line-of-sight (LOS) velocity, and microturbulence from imaging spectropolarimetric observations in CBGs observed in the \CaK{} line with the highest known spatial, spectral, and temporal resolution data.

The \CaK{} profiles in the CBGs show an enhanced \ktwov{} peak and a hint of redshift of the \kthree{} feature.
In the center of the CBG structure when it is at its maximum,  we find an average enhancement in temperature at \logtau{} = $-4.2$ of 1.1~kK and a maximum enhancement of up to $\sim4.5$~kK.
These enhancements in temperature are \mbox{colocated} with upflows in the LOS velocity between \logtau{} = [$-$4.2, $-$3].
%
The average strength of these upflows is $-$2.5~\kms{} and can be as large as $-$6~\kms{}.
However, the extreme values of temperature enhancements and upflows appear only in a few pixels in an acoustic shock region (see Fig.~\ref{fig:preshockpeakshock}).
%
Above \logtau{} = $-$4.2, we found strong downflows greater than $+8$~\kms{} which are stronger than the downflows observed in the quiescent atmospheres prior to the onset of the CBGs.
The retrieved value of microturbulence in the atmosphere of CBGs is negligible at chromospheric layers.
%
As explained in section~\ref{subsect:FOVanalysis}, there could be some degeneracy between the temperature and the microturbulence which may lead to overestimation of the values of the temperature enhancement.
However, following our experiment described in appendix~\ref{sect:tvturbrel}, we conclude that the uncertainty in temperature enhancement is not very significant.
%
Analysis of temperature response functions suggests that during the peak phase of the CBGs, the \CaK{} line is most sensitive to a perturbation in the temperature at \logtau{} = $-$4.2 in contrast to the quiescent atmosphere which is sensitive at about \logtau{} = $-$5.5.
We corroborate the findings by \citet{2020A&A...634A..56D}, who found that at the time of shock propagation, using simultaneous inversions of IRIS and ALMA data, the $\mathrm{k_{2V}}$ peak of the \ion{Mg}{II}~k line have a predominant contribution from the atmospheric layers that are sensitive around \logtau{} = $-$4.2.
%
%
During the CBG activity, we found temperature enhancements with the upflows in the lower chromosphere and downflows in the upper chromosphere in the direction of the LOS.
As the CBG progresses, the upper chromosphere (\logtau{}$>-$4.2) gets more and more downflowing and the lower chromosphere (\logtau{} = [$-$4.2, $-$3]) gets more and more upflowing.
The maximum response to perturbations in temperature and LOS velocity is at upflowing lower chromospheric layers, giving observational support to the interpretation that CBGs are manifestations of upward propagating acoustic shock waves in downflowing atmospheres \citep{1997ApJ...481..500C}.
The upflows in the lower chromosphere shift the opacity responsible for the two \ktwo{} features symmetrically positioned about the \CaK{} line core to the blue wing of the \CaK{} line.
These upward propagating shock waves enhance the gas density at lower chromospheric layers which couples the \ion{Ca}{ii} populations to the local conditions.
The local temperature enhancements at lower chromosphere result in an enhancement in the source function in the blue wing of the \CaK{} line (nominal \ktwov{} position).
Since the upper chromosphere is downflowing, the overlying opacity is redshifted, i.e., there is little opacity to absorb this blueshifted radiation, resulting in enhanced emission at the \ktwov{} wavelength position of the \CaK{} line \citep{1997ApJ...481..500C}.
The downflows above \logtau{} = $-$4.2 suggest that the \kthree{} opacity is shifted redward of line core which causes an opacity removal effect at the \ktwov{} peak, which enhances the \ktwov{} intensity while suppressing the \ktwor{} peak, an effect termed as ``Opacity Window'' by \citet{2019A&A...631L...5B, 2021A&A...654A..51B} who studied the formation of the \CaK{} line of the on-disk spicules.
However, response to perturbations in LOS velocity is only significant between \logtau{} = [$-$4.5, $-$3.5] and negligible at higher atmospheric layers, hence the amplitudes of these downflows are not reliable.
The above results underline the role of velocity gradients in the atmosphere in producing excess emission in one of the two (blue or red) wings of spectral lines.
Velocity gradients have also been found to be the cause of asymmetric emission in other physical mechanisms.
The effect of wave train of upflows and downflows on the line source function has been studied by \citet{1981ApJ...249..720S, 1984mrt..book..173S} who suggest that velocity gradients can enhance one of the \ktwo{} peaks.
\citet{2015ApJ...810..145D} found an enhancement in the red wing intensity of the \CaIR{} line due to velocity gradients produced by an upflowing magnetic bubble against a downflowing background.
Steep velocity gradients due to upflows in lower chromosphere and downflows in upper chromosphere can give rise to excess emission in blue wing (blue asymmetry) of the \Halpha{} spectral line and vice-versa \citep{2015ApJ...813..125K}.

The \CaK{} profiles in the atmosphere surrounding the core of the CBGs also show a relatively weak enhancement in the \ktwov{} peak intensity and redshifted \kthree{} features however we did not find upflows in LOS velocity in the lower chromosphere.
The average temperature enhancement seen in such atmospheres is 0.9~kK with maximum enhancement of 2~kK.
The enhancement in the \ktwov{} peaks of such profiles can be explained by the ``opacity window’’ effect where the \kthree{} opacity is redshifted, causing the removal of opacity at the \ktwov{} peak enhancing the feature \citep{2019A&A...631L...5B, 2021A&A...654A..51B}.

We also report that the temperature at \logtau{} = $-4.2$ during acoustic shocks has a near-constant value irrespective of the temperature before the onset of the shocks.
The result can also be interpreted in a way that the temperature enhancement is larger for the plasma regions with cooler quiescent temperature, which is similar to what \citet{2020ApJ...892...49H} found through the inversions of \CaIR{} spectra in the case of umbral flashes.

The temperature of CBGs at \logtau{} = $-$3 decreases with the evolution of the CBG with minimum temperature at the peak phase of the CBG.
The decrease in temperature is on average 0.75~kK.
It could be an expression of adiabatic cooling caused by expansion after a shock wave has passed through the region.
The velocity flows are typically directed outwards from the region's center, causing adiabatic expansion and thus cooling \citep{2004A&A...414.1121W}.
However, the temperature decrease could also be an artifact of the inversions:
The Tikhonov regularization is used to enforce smoothness in the model parameters, thus to keep the model smooth, a sudden increase in temperature at node position \logtau{} = $-$4.5, could be followed by a decrease in temperature at the node position \logtau{} = $-$3.
The signature of this temperature decrease is also reflected in the relatively poor fit of the \CaK{} line wings compared to the \ktwov{} emission feature (see appendix \ref{sect:qualityoffits}).
Moreover, the response to perturbations in temperature is much stronger in the higher layers (\logtau{} $\simeq -4.2$) compared to the lower atmospheric layers and hence this decrease in temperature is less reliable.

The temperature enhancement in the core of the CBGs ($\sim1$--$4.5$~kK), at spatial locations with maximum brightness in the bright grains, is up to five-fold compared to the enhancement reported by \citet{2013A&A...553A..73B} who studied the grains in the \CaH{} line assuming LTE conditions, and two-fold compared to the enhancement found by \citet{2020A&A...644A.152E} who studied such brightness temperature enhancements in the millimeter continuum using the data from ALMA.
The value of temperature enhancement is more reliable than the above works because,
a) LTE is not a valid approximation to model the \CaHK{} lines which form in the upper chromosphere and where PRD effects are significant \citep{1974ApJ...192..769M, 1989A&A...213..360U}, and
b) the ALMA observations studied by \citet{2020A&A...644A.152E} had a lower spatial resolution (2\arcsec) which is larger than the size of the CBGs we have studied in this paper. This may lead to reduced brightness temperatures and therefore an underestimation of the temperature enhancement as pointed out by \citet{2021A&A...656A..68E}.

Umbral flashes (UFs), which are interpreted as manifestations of magneto-acoustic shock waves, also show single peaked emission about nominal \ktwov{} position of the \CaK{} line with redshifted \kthree{} features and an enhancement in the blue wing of the \CaIR{} line with redshifted line core.
The Doppler shifts of the emitting flows in the UF atmospheres have been discussed by many authors in the literature with both upflowing \citep{2013A&A...556A.115D, 2018A&A...619A..63J, 2019ApJ...882..161A, 2020ApJ...892...49H} and downflowing \citep{2019A&A...627A..46B, 2017ApJ...845..102H, 1970SoPh...11..347A} chromospheres.
The above set of authors found an increase in temperature of about $1$--$2$ kK compared to the background which is comparable to our result.

Using a simulation of wave propagation in a sunspot umbra, \citet{2014ApJ...795....9F} showed that before the onset of the shocks the chromosphere is predominantly downflowing with typical absorption \CaIR{} profiles.
As the shock progresses, the middle chromosphere gets more and more upflowing whereas upper chromosphere gets more and more downflowing with an emission peak visible in blue wing of the \CaIR{} spectral line.
There is a cospatial temperature enhancement about the heights with upflowing plasma, which is about $1.5$~kK at the peak of the shock.
\citet{2020A&A...642A.215H} have reported counter-flowing solutions with weak upflows in the lower chromosphere and strong downflows in the upper chromosphere during UFs through the inversions of spectropolarimetric observations of the \CaIR{} line.
However, they argued that the magnitude of temperature enhancement can be overestimated due to the hydrostatic equilibrium assumption used in the inversion codes.
Using simulations it has been demonstrated that upward propagating shock waves increase the gas density at lower chromospheric layers, resulting in a strong coupling of the source function to the local conditions \citep{1997ApJ...481..500C}.
However, due to the requirement of the hydrostatic equilibrium assumption in inversions, such density perturbations cannot be modeled by any of the current inversion codes.
As a result, the density perturbations are ignored, and the temperature and the Doppler velocities are the only contributors to the source function enhancement, potentially leading to an overestimation of the magnitude of the temperature enhancement.

In summary, we report the evolution of temperature, LOS velocity and microturbulence from photosphere through lower to upper chromosphere through the simultaneous multi-line non-LTE inversions of spectroscopic observations of CBGs in the \CaK{}, \CaIR{} and \FeIline{} lines.
Our analysis of temperature and LOS velocity response functions support the interpretations using simulations \citep{1997ApJ...481..500C, 2004A&A...414.1121W} that CBGs are manifestations of upward propagating acoustic shocks against a background of downflowing atmospheres.

\begin{acknowledgements}
The authors thank the anonymous referee for the insightful comments.
HM thanks Jaime de la Cruz Rodríguez for technical support on STiC.
The Swedish 1-m Solar Telescope is operated on the island of La Palma by the Institute for Solar Physics of Stockholm University in the Spanish Observatorio del Roque de Los Muchachos of the Instituto de Astrofísica de Canarias.
The Institute for Solar Physics is supported by a grant for research infrastructures of national importance from the Swedish Research Council (registration number 2017-00625).
IRIS is a NASA small explorer mission developed and operated by LMSAL with mission operations executed at NASA Ames Research center and major contributions to downlink communications funded by ESA and the Norwegian Space Centre.
This research has made use of the High-Performance Computing (HPC) resources (NOVA cluster) made available by the Computer Center of the
Indian Institute of Astrophysics, Bangalore.
This research is supported by the Research Council of Norway, project numbers 250810, 
325491, 
and through its Centres of Excellence scheme, project number 262622.
S.B. gratefully acknowledges support from NASA contract NNG09FA40C (IRIS).
This research has made use of NASA’s Astrophysics Data System Bibliographic Services.
\end{acknowledgements}

\bibliography{report}   
\bibliographystyle{aa}   

\begin{appendix}

\section{$k$-means clustering}
\label{sect:kmeansclustering}
\begin{figure}[htbp]
    \centering
    \resizebox{0.5\textwidth}{!}{\includegraphics{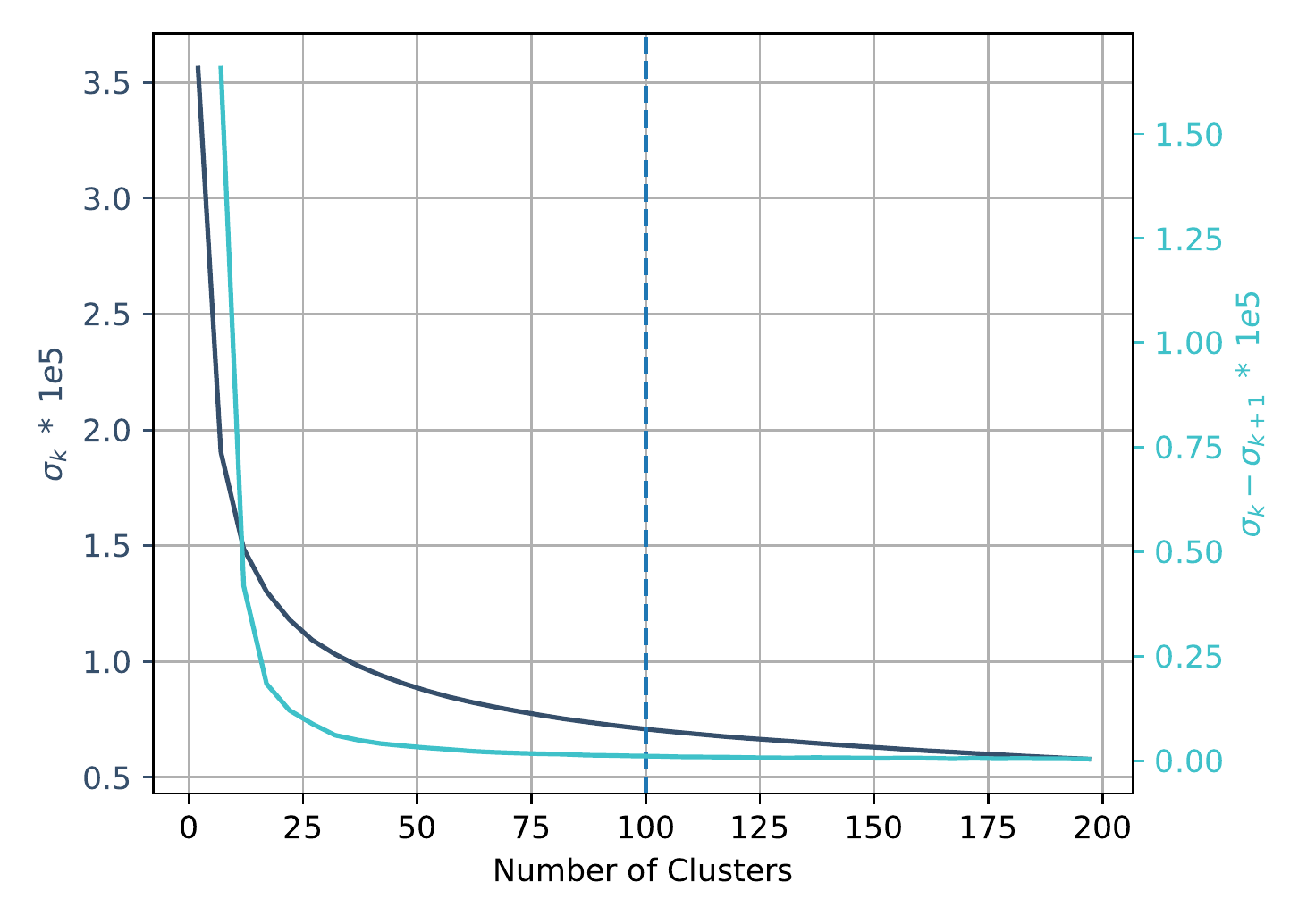}}
    \caption{Finding the optimum number of clusters $k$ for the $k$-means clusturing of \CaK{} profiles. The dark blue curve denotes inertia ($\sigma_{k}$) for k clusters, whereas $\sigma_{k} - \sigma_{k+1}$ is represented by cyan curve. The dashed vertical line indicates the used number of clusters $k=100$.}
    \label{fig:kmeansinertia}
\end{figure}

The $k$-means clustering \citep{zbMATH03340881} is one of the many unsupervised learning algorithms which are used to find patterns and structures in an unlabeled data set.
The algorithm works by partitioning an unlabeled data set ($m$ data points with $n$ features) into $k$ clusters.
In our case, the unlabeled data set are the spatially resolved image elements and features are 30 wavelength positions, 29 wavelength positions sampled in the \CaK{} line and 1 continuum 4000~\AA\, position.
The algorithm is iterative whose objective is to minimize inertia ($\sigma_{k}$), i.e., within cluster sum of squared distances from the cluster center.
A cluster center is nothing but the mean of all data points belonging to the cluster.
In our case, a cluster center is the mean of all spectral profiles belonging to the cluster, which we call Representative Profile (RP).
The algorithm is initialized by $k$ number of predefined cluster centers and each data point is assigned to a cluster with the nearest (euclidean distance, $\chi^{2}$) center.
In subsequent iterations, new cluster centers are calculated from the clusters defined in previous iteration and the process continues until the algorithm converges.
The performance of $k$-means clustering algorithm have a heavy dependence on the seed cluster centers.
Random selection of seed cluster centers generally result in a poor clustering.
Thus, to improve the quality of clustering, we used $k$-means++ \citep{10.5555/1283383.1283494} algorithm which ensures that seed cluster centers are as far away from each other as possible.
The $k$-means++ algorithm proceeds in an iterative manner by first selecting a cluster center at random, then each new center (up to $k$) is chosen such that the distance from the nearest previously chosen center is maximum.

Our main objective of performing $k$-means algorithm is to identify CBG like profiles and use the stratification of atmospheric parameters inferred from inversions of RPs as initial guess atmosphere while inverting actual observations.
Hence, the actual spectral profiles must be close to the corresponding RPs at all wavelength positions.
The $k$-means clustering algorithm calculates the euclidean distance between a data point and the cluster mean before assigning it to a cluster which implies that while finding the nearest cluster, the $\chi^{2}$ is affected more for the wavelength positions that have higher variance.
Hence, we must normalize the data before feeding it to the $k$-means algorithm which is done by subtracting the mean and dividing by standard deviation at each wavelength position before doing $k$-means clustering, ensuring that all wavelength positions have a variance of one.
However, we have more wavelength samples between the \CaK{} far wings and the \kone{} features (19) compared to the samples between the two \kone{} features (10).
In addition, we have one wavelangth point in the 4000~\AA\, continuum too.
Since, the radiation at wavelength positions between the \kone{} features originate from higher atmospheric layers (lower and upper chromosphere) compared to that of at wavelength positions beyond the \kone{} features and the continuum position, it will adversely affect the atmospheric parameters retrieved from inversions of RPs at chromospheric layers, which is detrimental to our study.
Therefore, we multiplied every pixel with a normalized weight such that the sum of variances at wavelength positions between the \kone{} features equals to the of sum of variances at wavelength positions beyond the \kone{} features and the continuum point.

Thereafter, seven time-frames under the best seeing conditions were selected for clustering using the $k$-means algorithm, which contained a total of 15,988,896 pixels.
We used elbow method to determine the optimal number of clusters $k$ in which our dataset can be grouped, where we analyze change in $\sigma_{k}$ with respect to $k$.
we performed clustering of the data with $2 \leq k \leq 197$ and plotted the $\sigma_{k}$, and also the change ($\sigma_{k} - \sigma_{k+1}$) as shown in Elbow plot, Fig.~\ref{fig:kmeansinertia}.
The $k$ is chosen such that $\sigma_{k}$ decreases linearly.
However, in reality, it is generally a smooth curve, and hence we choose a $k$ large enough that $\sigma_{k} - \sigma_{k+1}$ does not change.
According to the Elbow plot, the $\sigma_{k} - \sigma_{k+1}$ does not change after $k\sim70$, however our purpose of $k$-means is to use the guess atmospheres inferred from inversions of the RPs as an initial guess in inversions of the actual observations, we chose a larger value $k=100$, to get a better guess atmospheric model.

\section{Quality of fits}
\label{sect:qualityoffits}
\begin{figure*}[htbp]
    \centering
    \resizebox{0.19\textwidth}{!}{\includegraphics[]{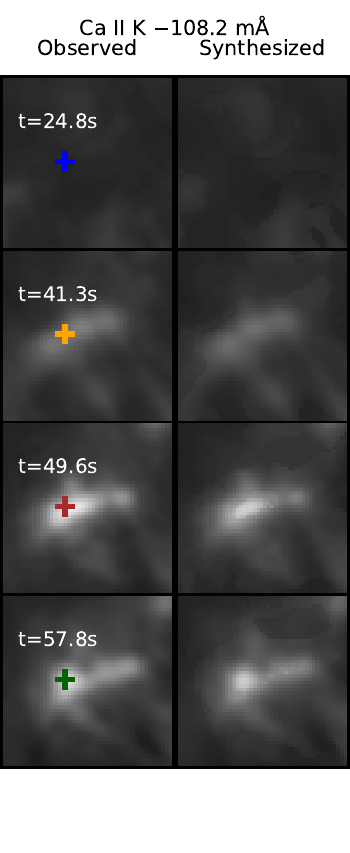}}
    \resizebox{0.19\textwidth}{!}{\includegraphics[]{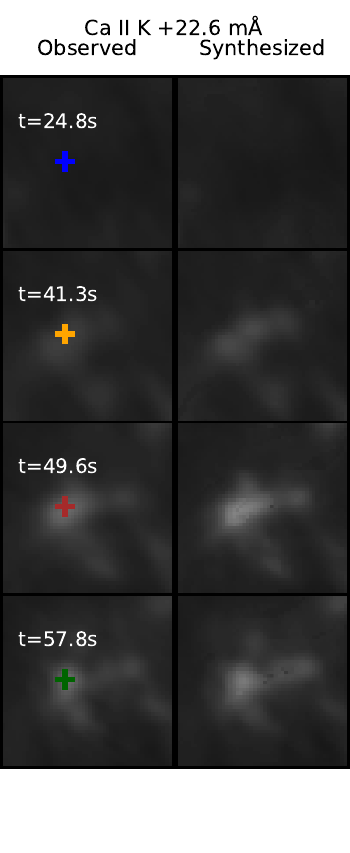}}
    \resizebox{0.19\textwidth}{!}{\includegraphics[]{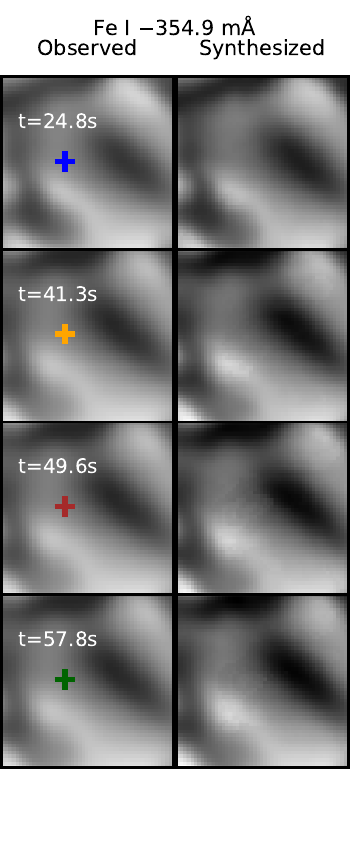}}
    \resizebox{0.31\textwidth}{!}{\includegraphics[]{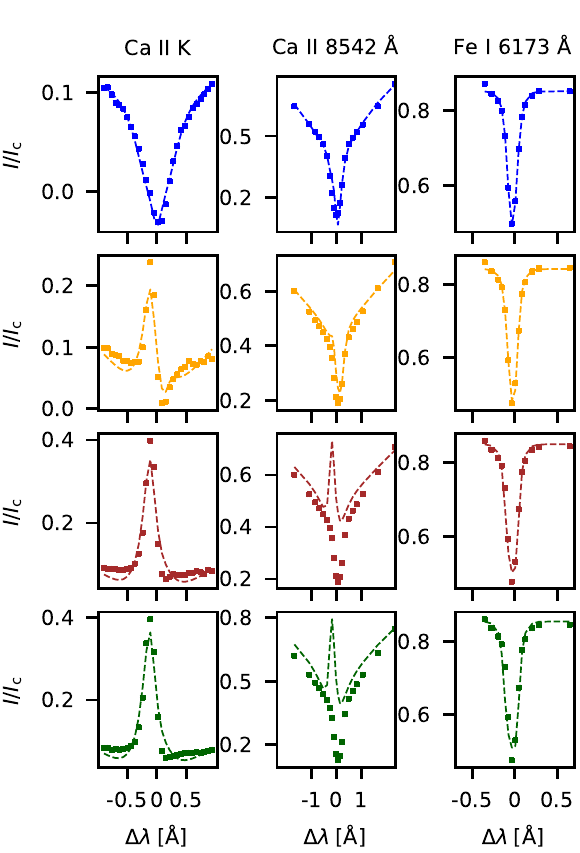}}
    \caption{Illustrating the quality of fits of synthesize narrowband images and spectral profiles inferred using inversions for \roi{A}. Comparison of synthesized with observed narrowband images at wavelength offsets of $-108.2$~\mAA and $+22.6$~\mAA from the \CaK{} line core and $-$354.9~\mAA{} from the \FeIline{} line core at different times. A gamma correction is applied with $\gamma=0.7$ on both observed and synthesized \CaK{} narrowband images before display. The right panel shows the evolution of the observed (dotted) and synthesized (dashed) profiles of a pixel that appear brightest at $t=49.6$~s in the narrowband images at wavelength offset of $-108.2$~\mAA from the \CaK{} line core. The \CaIR{} and the \FeIline{} spectral cubes were acquired 27~s before the \CaK{} cube.}
    \label{fig:qualityoffits}
\end{figure*}

\begin{figure*}[htbp]
    \centering
    \resizebox{0.19\textwidth}{!}{\includegraphics[]{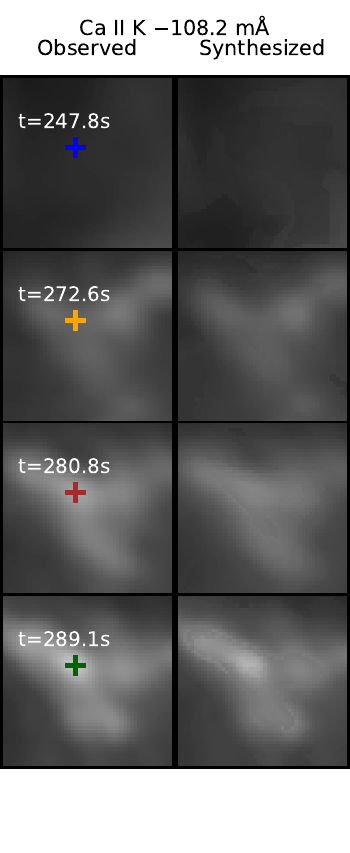}}
    \resizebox{0.19\textwidth}{!}{\includegraphics[]{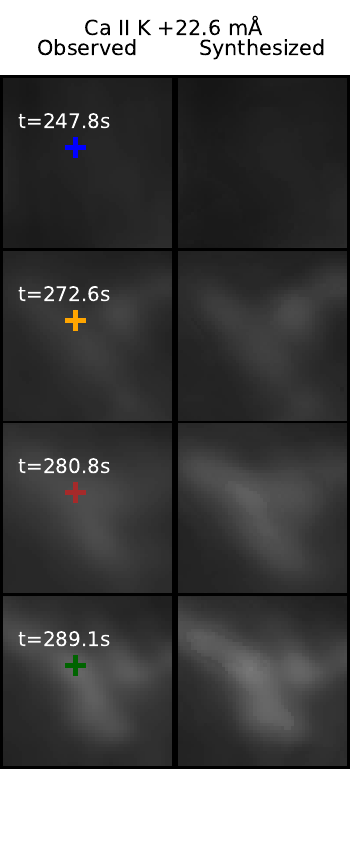}}
    \resizebox{0.19\textwidth}{!}{\includegraphics[]{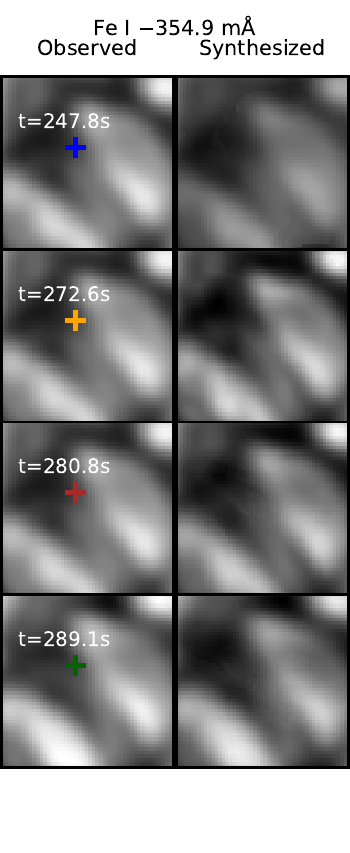}}
    \resizebox{0.31\textwidth}{!}{\includegraphics[]{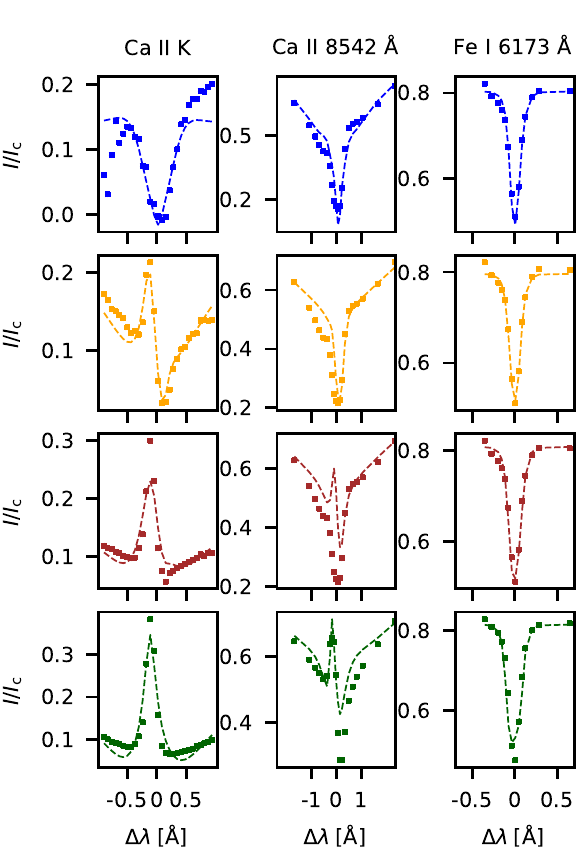}}
    \caption{Plots illustrating the quality of fits in \roi{B} in the same format as Fig.~\ref{fig:qualityoffits}. The \CaIR{} and the \FeIline{} spectral cubes were acquired 8~s before the \CaK{} cube.}
    \label{fig:qualityoffits_roib}
\end{figure*}
We discuss the match between synthetic and observed Stokes $I$ profiles in the \CaK{}, \CaIR{} and \FeIline{} lines for \roi{A} and B in Fig.~\ref{fig:qualityoffits} and \ref{fig:qualityoffits_roib} .

The overall morphological structure of CBG is very well reproduced in the narrowband images of the \CaK{} line.
A typical granulation structure can be seen in the synthesized \FeIline{} narrowband images, closely resembling to that of in observed images.
The synthesized profiles of the \CaK{}, \CaIR{} and \FeIline{} lines show a good fit with the observed ones with relatively better fitting of the \ktwov{} emission feature compared to the \kone{} features and the line wing wavelength positions.
The synthesized \CaIR{} profile of the CBG in \roi{A} is in emission compared to the observed profile, which shows typical absorption because the evolution of the CBG is poorly captured in the \CaIR{} data due to CRISP having $\sim4.5$ times less cadence than the CHROMIS.
However, in \roi{B} the observed \CaIR{} spectral line is seen in emission, suggesting that the observation coincided with the maximum phase of the CBG seen in the \CaK{}, and qualitatively similar to the synthesized \CaIR{} spectra.
This suggests although the \CaIR{} line has one-fourth the weightage in inversions compared to the \CaK{}, if the observation in the \CaIR{} coincides with the CBG, we are able to get good fits with the utilized weighing scheme.

\section{Relationship between \temp{} and \vturb{}}
\label{sect:tvturbrel}
\begin{figure*}
    \includegraphics[width=0.5\textwidth]{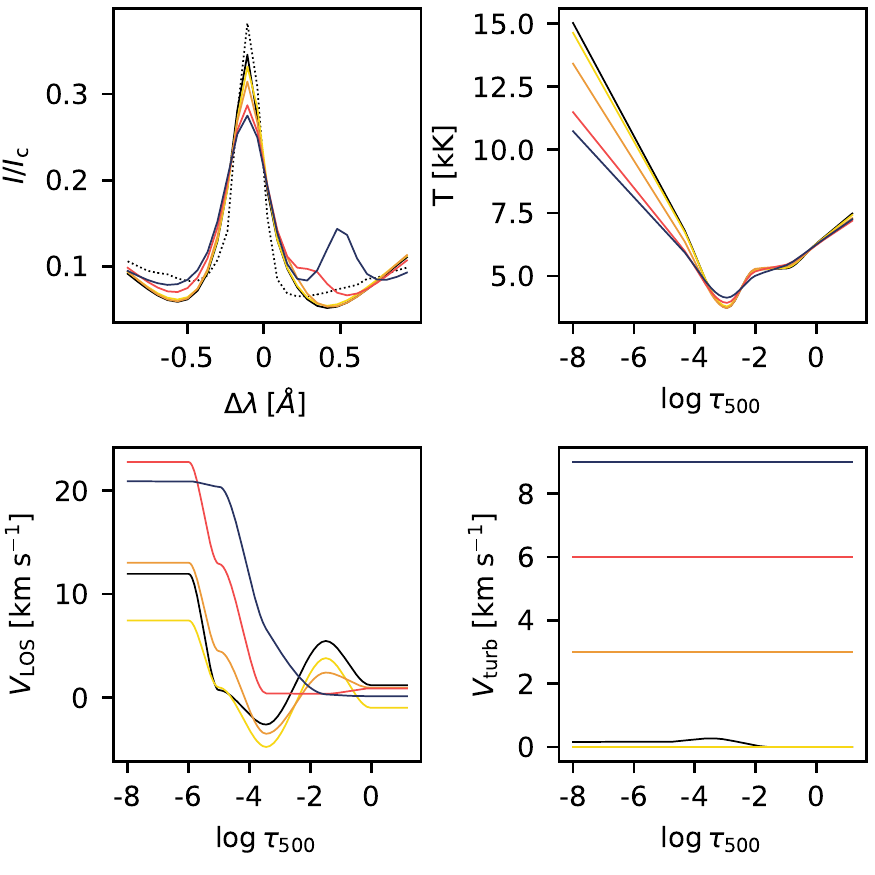}
    \includegraphics[width=0.5\textwidth]{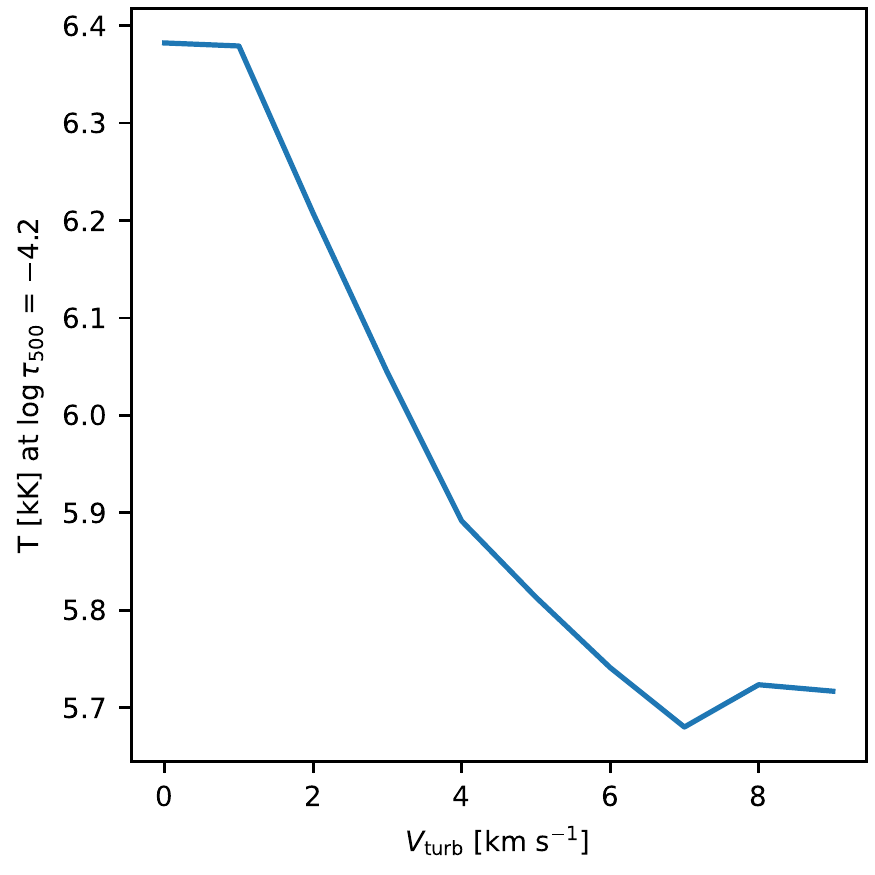}
    \caption{Understanding the relationship of \temp{} and \vturb{}. The top-left panel shows the quality of fits to the \CaK{} profile, the panel to the right shows the \temp{} stratification. The bottom left panel shows the \vlos{} stratification and the panel to the right shows the value of the \vturb{} used to invert the selected profile. The large panel at right shows the \temp{} at \logtau{}=$-$4.2 as a function of \vturb{}. The observed (dashed) and fitted (solid) \CaK{} profile (the \CaK{} profile shown in right panel of Fig.~\ref{fig:qualityoffits_roib} at $t$ = 57.8~s) is shown in black color. The \temp{}, \vlos{} and \vturb{} stratification inferred using inversions (main text, \vturb{}$\simeq$0) are also shown in black color. The yellow, orange, red and blue colors represent the experiments for the \vturb{} values 0, 3, 6 and 9~\kms{} respectively.}
    \label{fig:tvturb}
\end{figure*}

To estimate the uncertainty in the enhancement of \temp{} due to the degeneracy between \temp{} and \vturb{}, we performed an experiment where we inverted the brightest pixel in \roi{B} with different values of \vturb{}. 
%
The maximum value of \vturb{} inferred from the quiescent pixels is found to be about 5~\kms{}.
Therefore, we performed 10 inversions in total with fixed \vturb{} values (ranging from 0--9~\kms{}) per inversion, but nodes in \temp{} and \vlos{} were placed in the same way as described in Table~\ref{table:inversionstrategy}.
The variation of \temp{} with \vturb{} is shown in Fig.~\ref{fig:tvturb}.
We find that for 1~\kms{} increase in the \vturb{}, the \temp{} at \logtau{} = $-$4.2 decreases roughly by 125~K.
However, the increase in the value of \vturb{} is correlated with a decrease in the \ktwov{} intensity which further leads to a worsening of the fit.
%
%
For example, the red and blue colored fitted profiles (with \vturb{} values of 6 and 9~\kms{}) have the least intensity at the \ktwov{} wavelength position.
In addition, the \ktwor{} appears as an emission peak which is not a characteristic of the observed grain profiles.
Thus, we come to the conclusion that minimal values of \vturb{} are required to achieve satisfactory fits of the observed \CaK{} profiles and that the uncertainty in the enhancement of the \temp{} as a result of low \vturb{} is not very significant.

\section{Supplementary figures}
\label{sect:supplementary}

This section shows supplementary figures for the figures shown in the main body.

\begin{figure*}
\centering
\resizebox{0.18\textwidth}{!}{\includegraphics{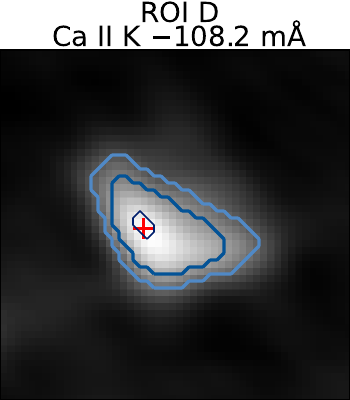}}
\resizebox{0.18\textwidth}{!}{\includegraphics{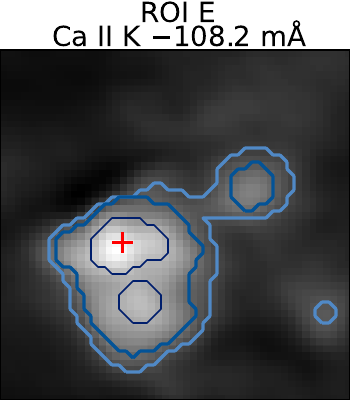}}
\resizebox{0.18\textwidth}{!}{\includegraphics{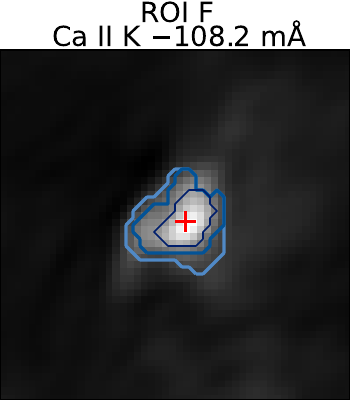}}
\resizebox{0.18\textwidth}{!}{\includegraphics{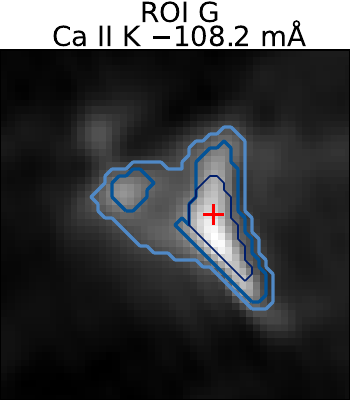}}
\resizebox{0.18\textwidth}{!}{\includegraphics{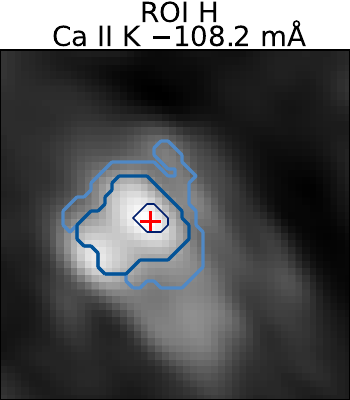}}
\resizebox{0.18\textwidth}{!}{\includegraphics{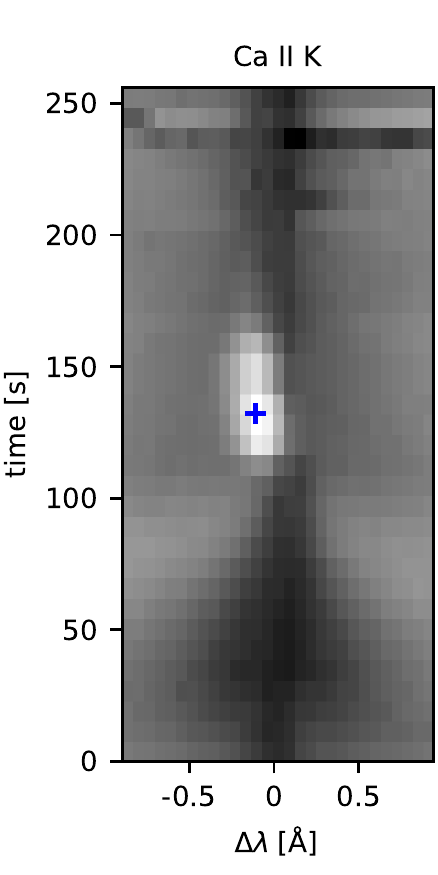}}
\resizebox{0.18\textwidth}{!}{\includegraphics{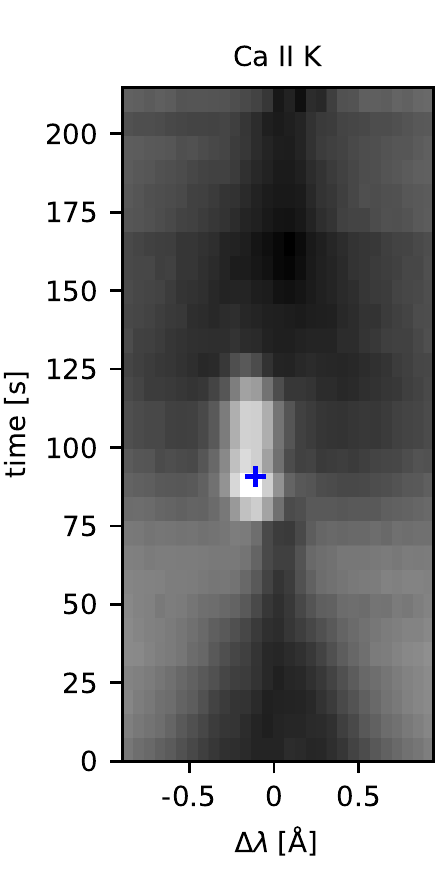}}
\resizebox{0.18\textwidth}{!}{\includegraphics{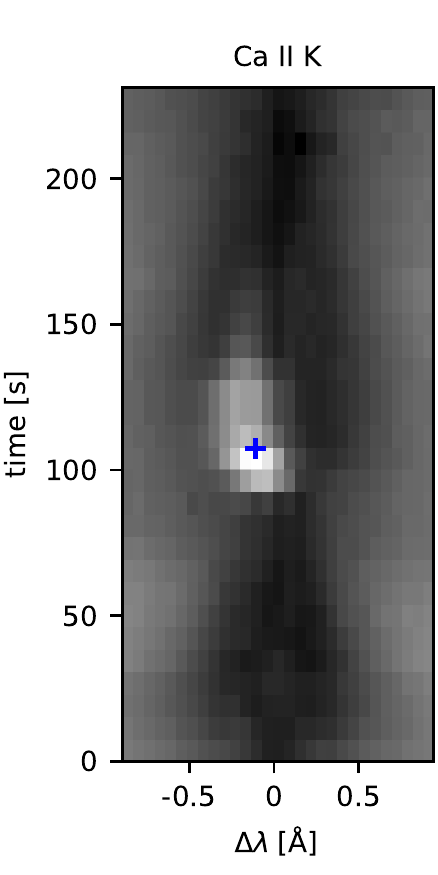}}
\resizebox{0.18\textwidth}{!}{\includegraphics{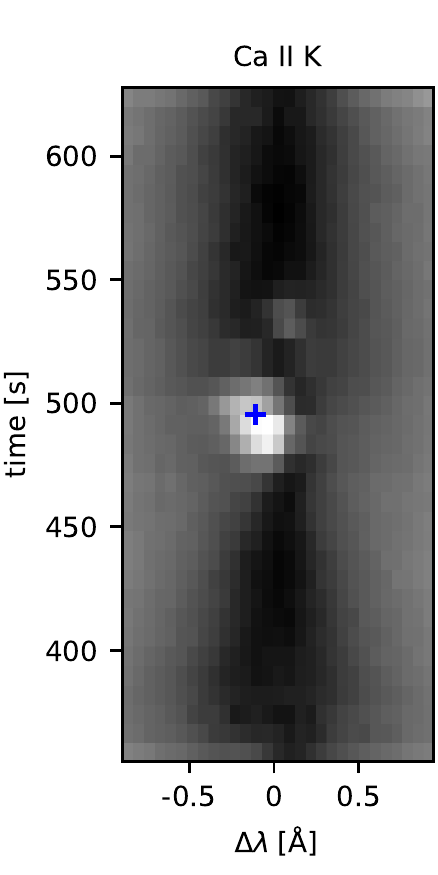}}
\resizebox{0.18\textwidth}{!}{\includegraphics{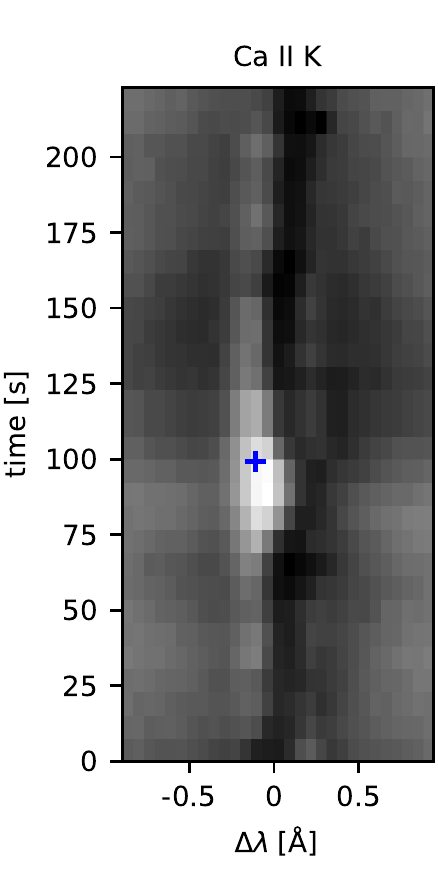}}
\caption{
An overview of the CBGs in ROI D--H. Panels in the top row show narrowband images at a wavelength offset of $-108.2$~\mAA{} from the \CaK{} line of ROI D-H. Panels in the bottom row show the \lambdat{} diagram in the \CaK{} line for the pixel marked with '+' in the top row. The time step at which narrowband images are shown in top row are marked with '+' in the bottom row.
}
    \label{fig:supplemtFOV1}
\end{figure*}

The evolution of the CBGs in the remaining five ROIs (D--H) is similar to the CBGs described in the main text (see Fig.~\ref{fig:supplemtFOV1}).
Before the onset of the CBG, the \CaK{} line shows typical absorption without any \ktwo{} spectral features (see \lambdat{} diagram).
At the onset of the CBG, there is an enhancement in the \ktwov{} peak intensity which is evident in the \lambdat{} diagram.
CBGs in \roi{E} and G show multiple sub-structures with multiple islands of intensity enhancements within the structure of the CBG.
The lifetime of the CBGs in \roi{D--H} varies from $\sim 30$~s to $\sim 60$~s.
However, the \ktwov{} spectral feature is present in the CBG in \roi{H} for a much longer duration.

\begin{figure*}[htbp]
\begin{center}
    \textbf{ROI D}
\end{center}
\begin{subfigure}{0.5\textwidth}
\resizebox{0.96\textwidth}{!}{\includegraphics[trim={0 {0.03\textwidth} 0 0},clip]{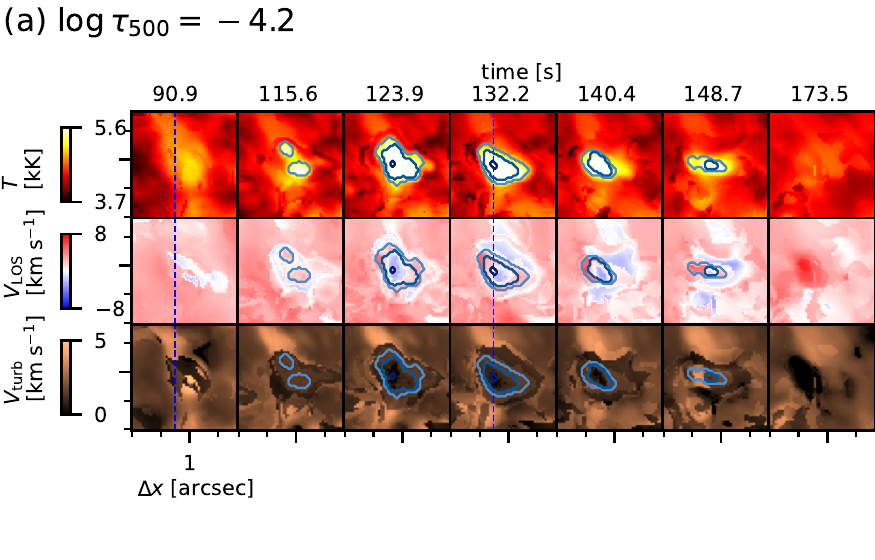}}
\resizebox{0.96\textwidth}{!}{\includegraphics[trim={0 {0.03\textwidth} 0 0},clip]{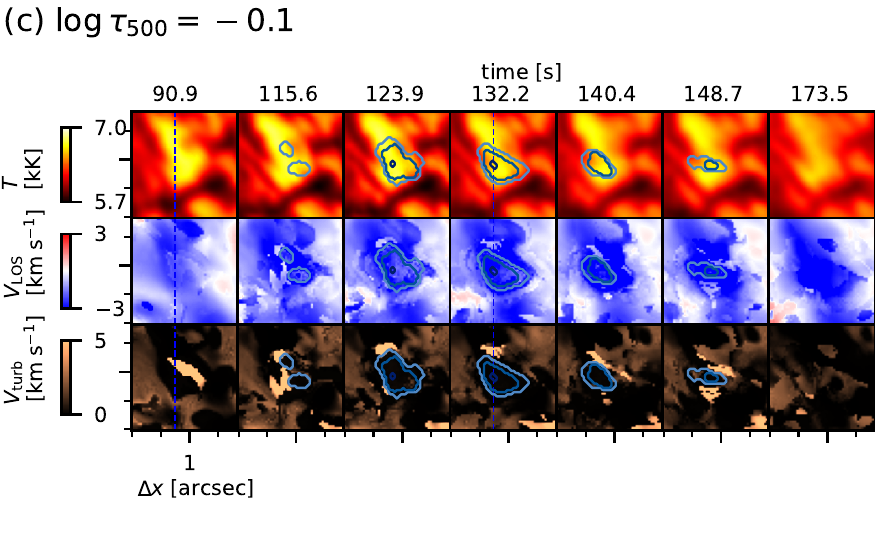}}
\end{subfigure}
\begin{subfigure}{0.5\textwidth}
\resizebox{0.96\textwidth}{!}{\includegraphics[trim={0 {0.03\textwidth} 0 0},clip]{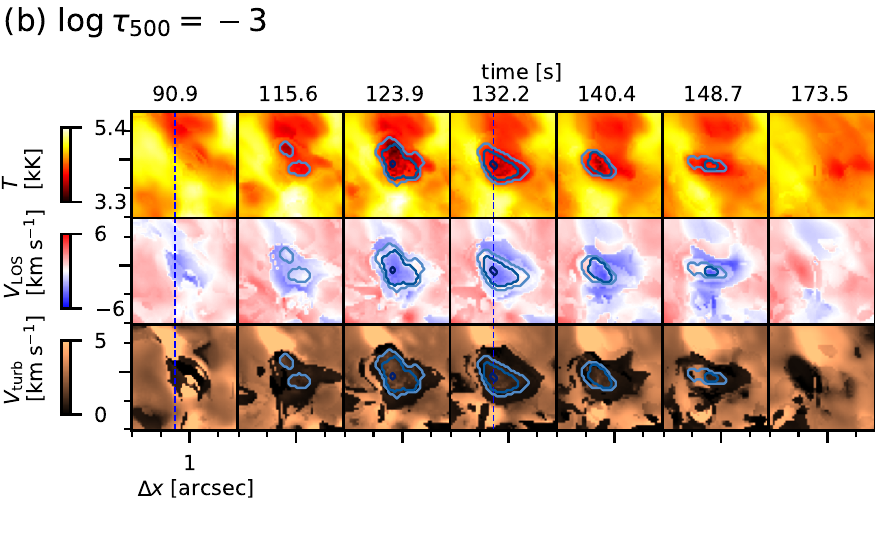}}
\resizebox{0.96\textwidth}{!}{\includegraphics[trim={0 {0.03\textwidth} 0 0},clip]{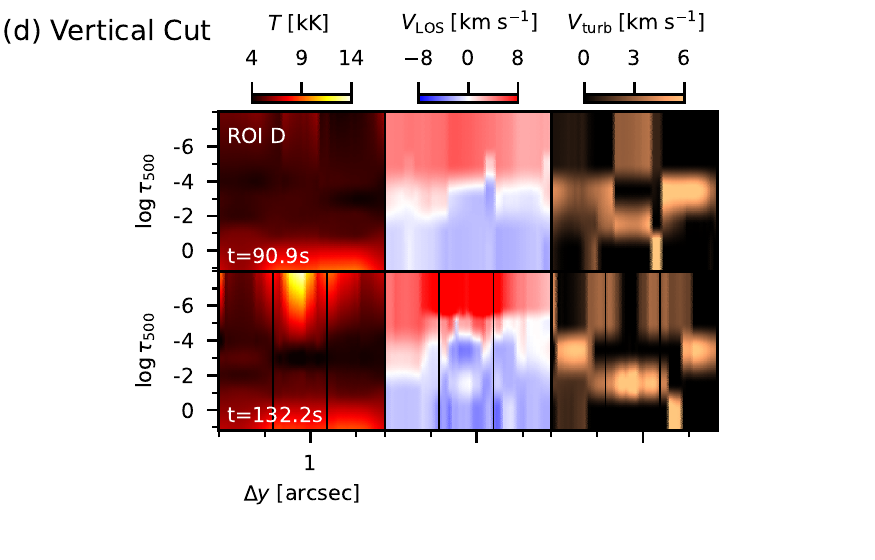}}
\end{subfigure}
\caption{Inversion results for \roi{D} in the same format as Fig.~\ref{fig:inversionmapa}.}
\label{fig:inversionmapd}
\end{figure*}

\begin{figure*}[htbp]
\begin{center}
    \textbf{ROI E}
\end{center}
\begin{subfigure}{0.5\textwidth}
\resizebox{0.96\textwidth}{!}{\includegraphics[trim={0 {0.03\textwidth} 0 0},clip]{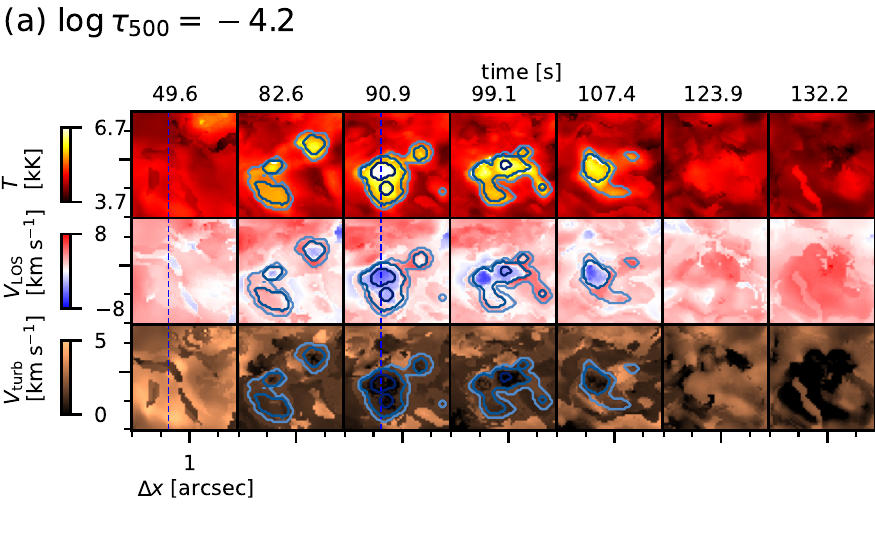}}
\resizebox{0.96\textwidth}{!}{\includegraphics[trim={0 {0.03\textwidth} 0 0},clip]{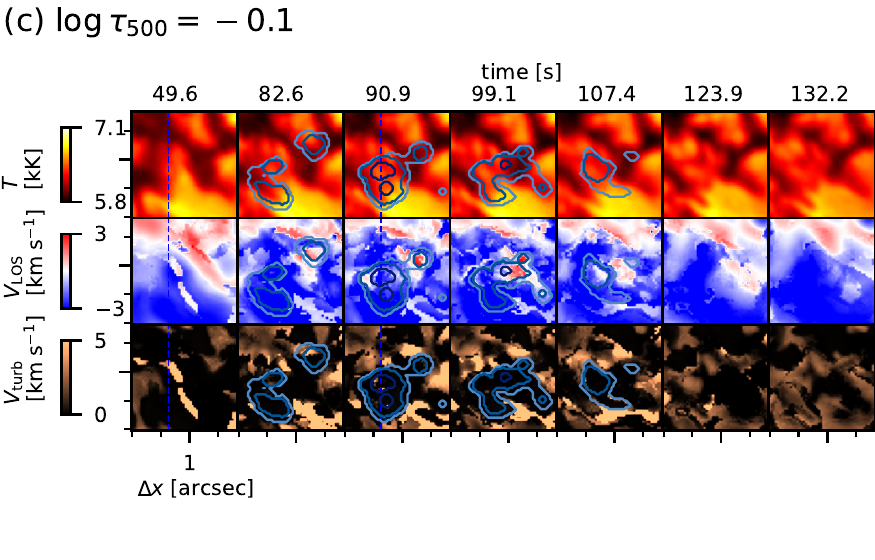}}
\end{subfigure}
\begin{subfigure}{0.5\textwidth}
\resizebox{0.96\textwidth}{!}{\includegraphics[trim={0 {0.03\textwidth} 0 0},clip]{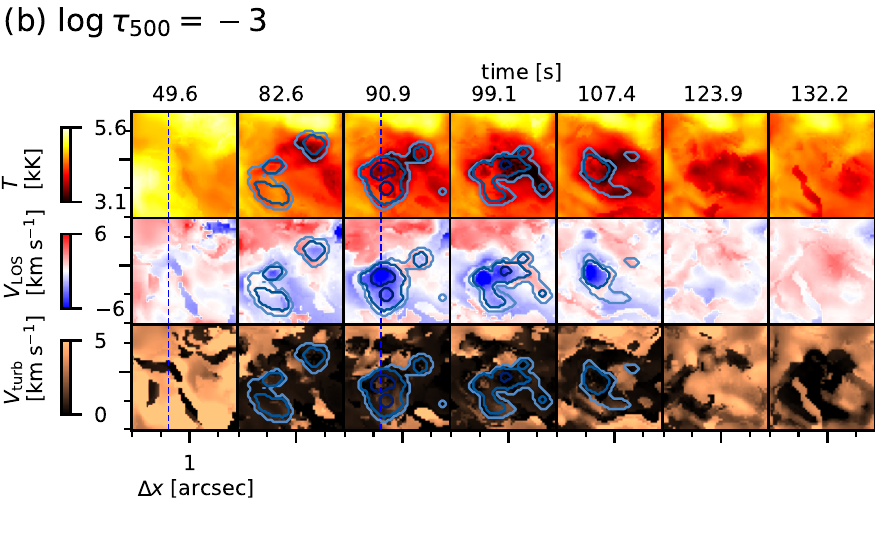}}
\resizebox{0.96\textwidth}{!}{\includegraphics[trim={0 {0.03\textwidth} 0 0},clip]{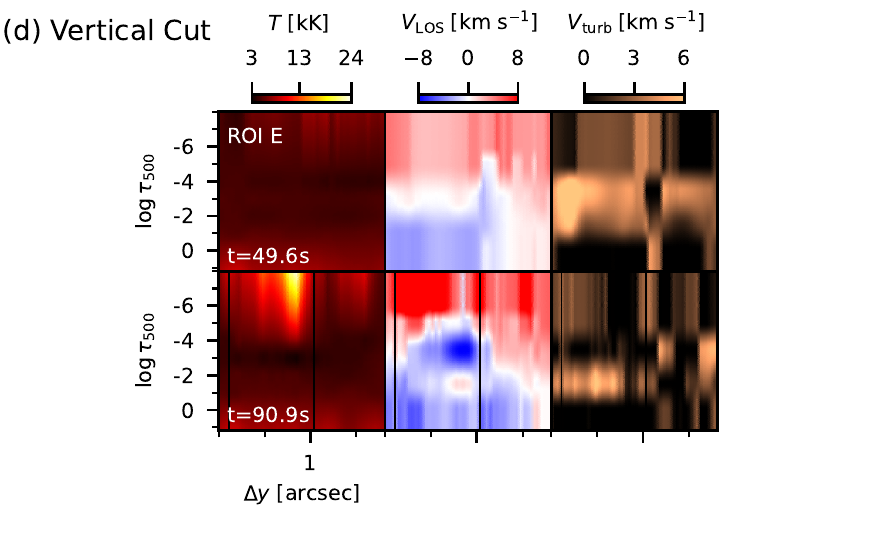}}
\end{subfigure}
\caption{Inversion results for \roi{E} in the same format as Fig.~\ref{fig:inversionmapa}.}
\label{fig:inversionmape}
\end{figure*}

\begin{figure*}[htbp]
\begin{center}
    \textbf{ROI F}
\end{center}
\begin{subfigure}{0.5\textwidth}
\resizebox{0.96\textwidth}{!}{\includegraphics[trim={0 {0.03\textwidth} 0 0},clip]{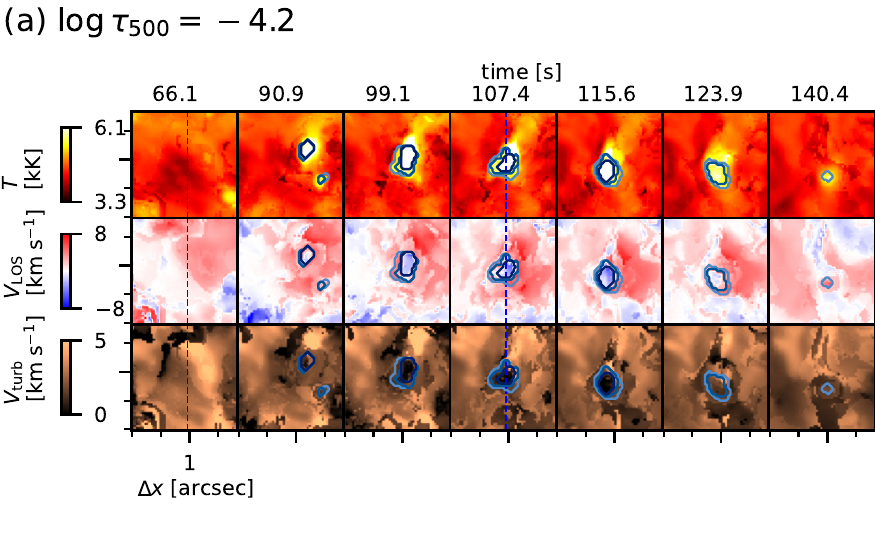}}
\resizebox{0.96\textwidth}{!}{\includegraphics[trim={0 {0.03\textwidth} 0 0},clip]{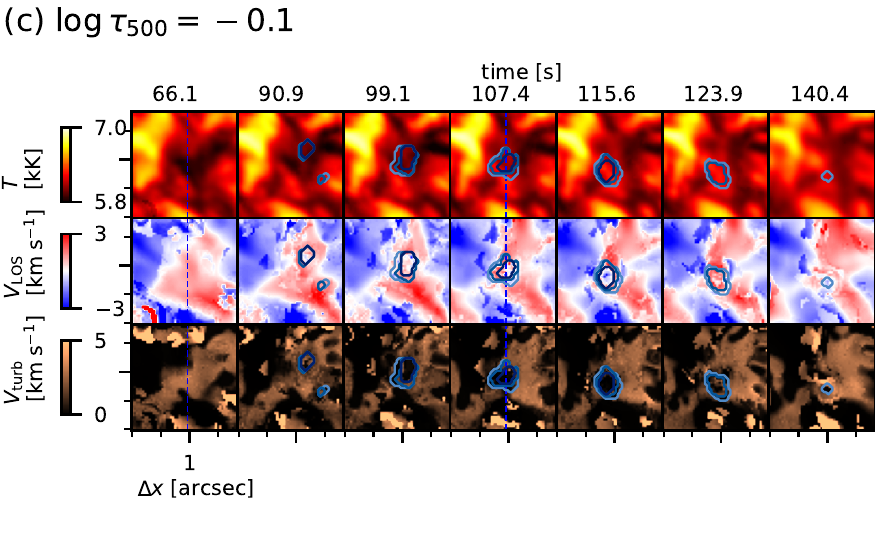}}
\end{subfigure}
\begin{subfigure}{0.5\textwidth}
\resizebox{0.96\textwidth}{!}{\includegraphics[trim={0 {0.03\textwidth} 0 0},clip]{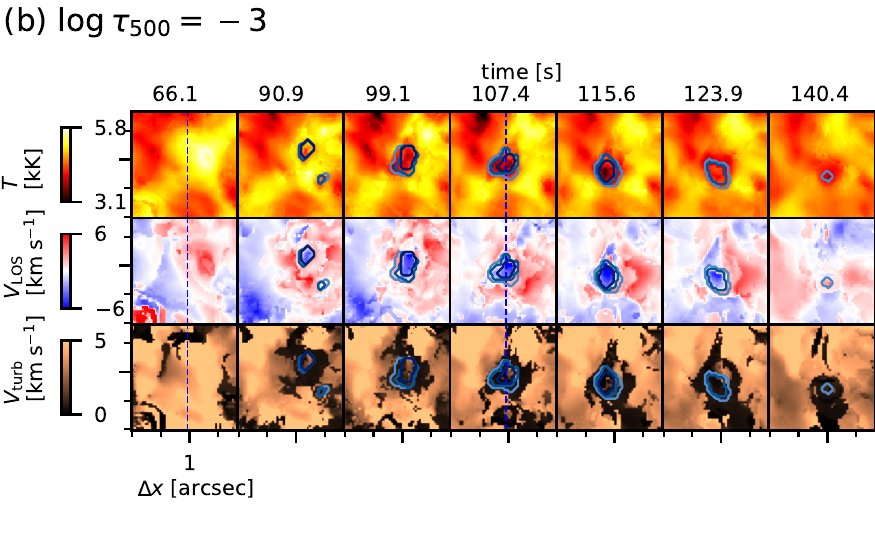}}
\resizebox{0.96\textwidth}{!}{\includegraphics[trim={0 {0.03\textwidth} 0 0},clip]{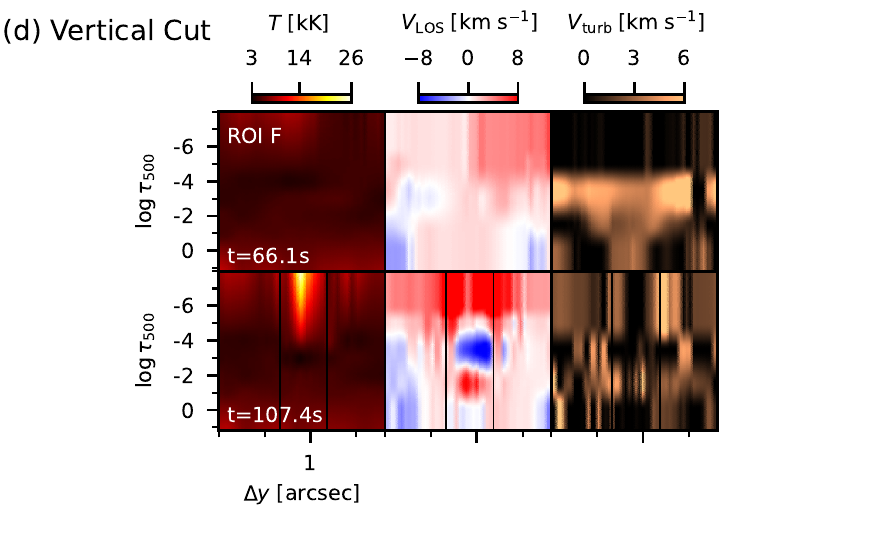}}
\end{subfigure}
\caption{Inversion results for \roi{F} in the same format as Fig.~\ref{fig:inversionmapa}.}
\label{fig:inversionmapf}
\end{figure*}

\begin{figure*}[htbp]
\begin{center}
    \textbf{ROI G}
\end{center}
\begin{subfigure}{0.5\textwidth}
\resizebox{0.96\textwidth}{!}{\includegraphics[trim={0 {0.03\textwidth} 0 0},clip]{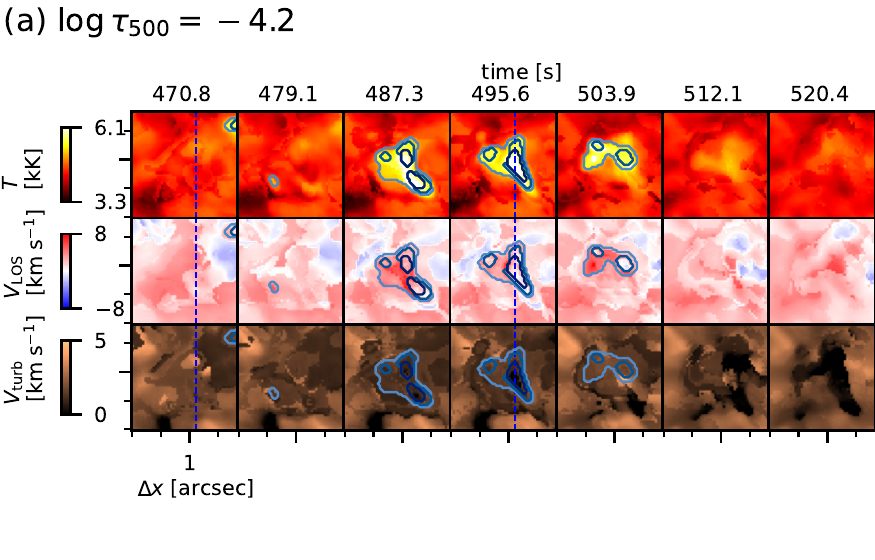}}
\resizebox{0.96\textwidth}{!}{\includegraphics[trim={0 {0.03\textwidth} 0 0},clip]{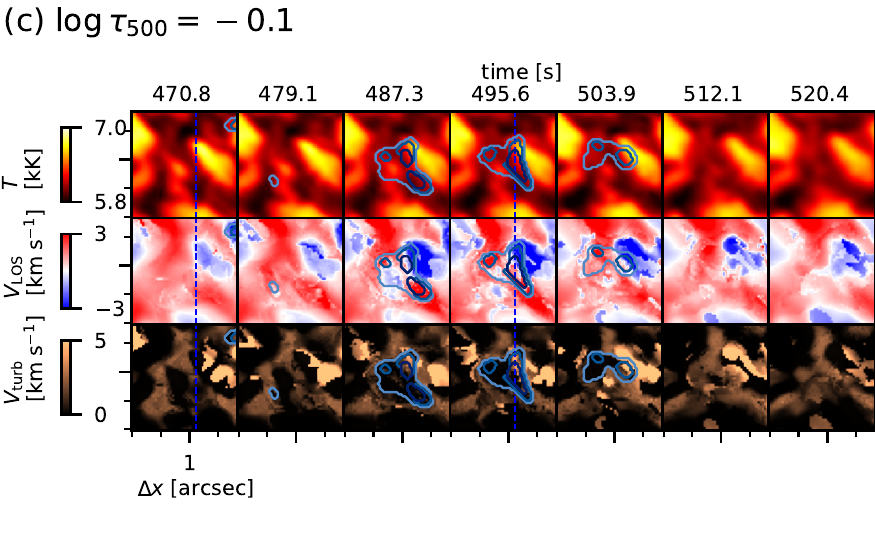}}
\end{subfigure}
\begin{subfigure}{0.5\textwidth}
\resizebox{0.96\textwidth}{!}{\includegraphics[trim={0 {0.03\textwidth} 0 0},clip]{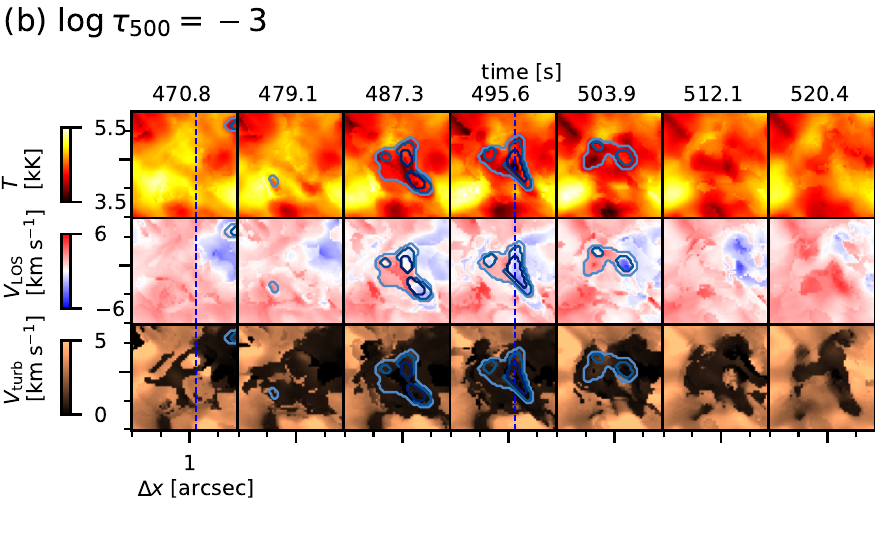}}
\resizebox{0.96\textwidth}{!}{\includegraphics[trim={0 {0.03\textwidth} 0 0},clip]{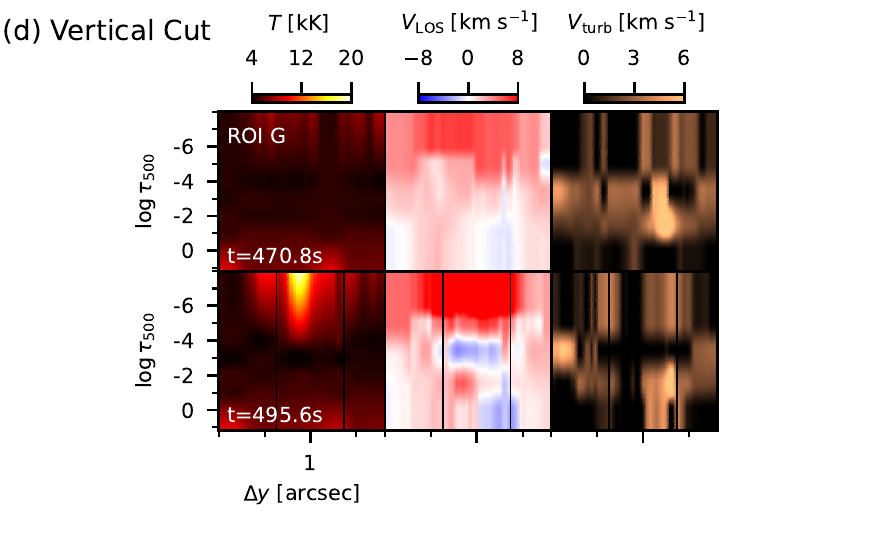}}
\end{subfigure}
\caption{Inversion results for \roi{G} in the same format as Fig.~\ref{fig:inversionmapa}.}
\label{fig:inversionmapg}
\end{figure*}

\begin{figure*}[htbp]
\begin{center}
    \textbf{ROI H}
\end{center}
\begin{subfigure}{0.5\textwidth}
\resizebox{0.96\textwidth}{!}{\includegraphics[trim={0 {0.03\textwidth} 0 0},clip]{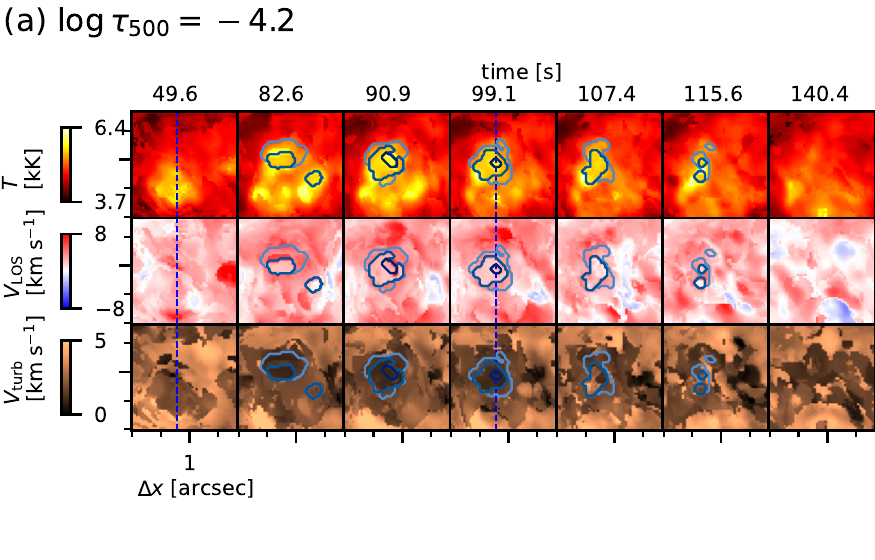}}
\resizebox{0.96\textwidth}{!}{\includegraphics[trim={0 {0.03\textwidth} 0 0},clip]{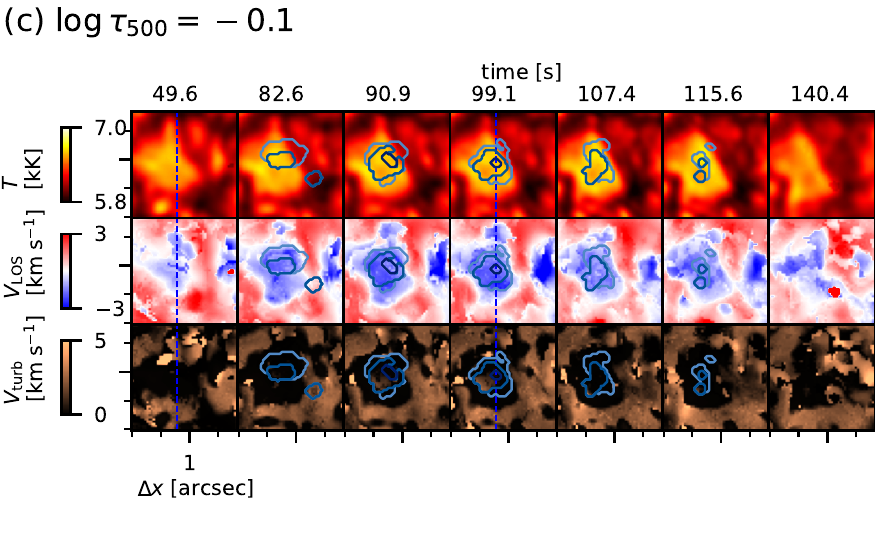}}
\end{subfigure}
\begin{subfigure}{0.5\textwidth}
\resizebox{0.96\textwidth}{!}{\includegraphics[trim={0 {0.03\textwidth} 0 0},clip]{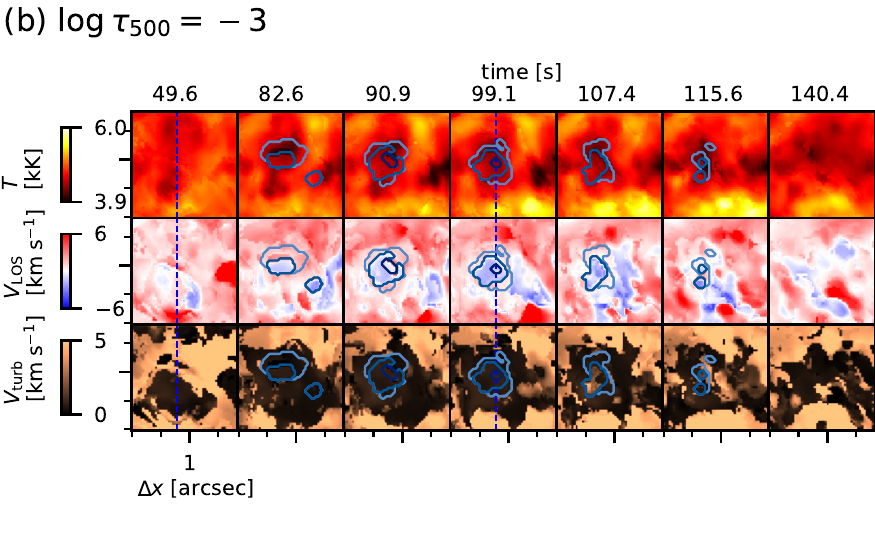}}
\resizebox{0.96\textwidth}{!}{\includegraphics[trim={0 {0.03\textwidth} 0 0},clip]{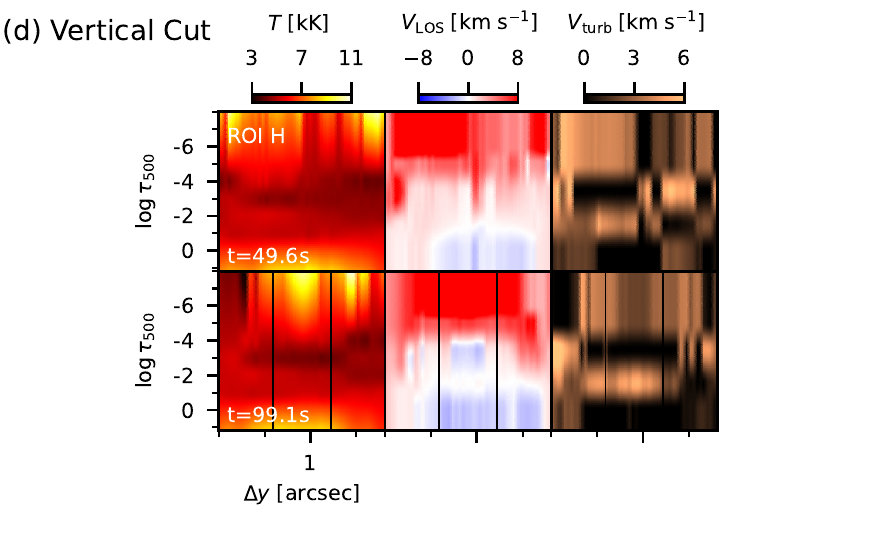}}
\end{subfigure}
\caption{Inversion results for \roi{H} in the same format as Fig.~\ref{fig:inversionmapa}.}
\label{fig:inversionmaph}
\end{figure*}

We present the evolution of the retrieved atmospheric parameters for the \roi{D--H} in Fig.~\ref{fig:inversionmapd}--\ref{fig:inversionmaph}, respectively.

The inferred atmospheric parameters of CBG in \roi{D}, E, and F evolve similar to the CBG in \roi{A} and B.
Before the onset of the CBG, at \logtau{} = $-4.2$, there is a little spatial variation in \temp{} with retrieved \vlos{} showing downflows of $\sim +4$~\kms{}.
As the CBG progresses, its area and intensity increase, there is a sharp enhancement in \temp{} of about 1.5--3~kK at the contours belonging to CBGs, with \vlos{} showing upflows up to $-6$~\kms{}.
The downflows observed in the upper chromosphere at \logtau{}$<-$4.2 have been enhanced with values greater than $+$8~\kms{}.
The \temp{} reaches maximum and \vlos{} reach minimum near-simultaneously before going back to the values before the onset.
Similar to \roi{A} and B, there is less microturbulence compared to pixels outside the CBGs.
The \temp{} at \logtau{} = $-3$ reaches minimum as the CBG progresses with \vlos{} showing upflows typically higher than that of at \logtau{} = $-4.2$.
The upflows in the vertical cut are colocated with the start of enhancement in \temp{}.
A typical granulation pattern is visible in photospheric layers, \logtau{} = $-$0.1.

The evolution of atmospheric parameters for CBG in \roi{G} and H is similar to \roi{C}.
There are no signatures of upflows at \logtau{} = $-4.2$ when the CBG is at its peak.
The retrieved \vlos{} at \logtau{} = $-3$ does not have a consistent structure, some pixels showing upflows and some downflows.
The upflows in \roi{G} and H at the region belonging to CBGs are located at an optical depth of about \logtau{} = $-4$ where the enhancement in \temp{} can be seen (see panel (d)).
The upflows and enhancement in \temp{} in \roi{H} are weaker than that of in \roi{G} (see panel (d)).

\end{appendix}
\end{document}